\documentclass[12pt]{report}
\usepackage[cp1251]{inputenc} % при необходимости для старых кодировок
\usepackage[english]{babel}
\usepackage{amssymb,amsmath}
\numberwithin{equation}{section}
\usepackage{graphicx}
\usepackage{tocloft}          % управление внешним видом оглавления
\usepackage{wrapfig} % <-- Добавляем этот пакет для обволакивания текстом
\usepackage{tikz}
\makeatletter
\renewcommand{\@makechapterhead}[1]{%
	\vspace*{0pt}%
	{\parindent \z@ \raggedright
		\ifnum \c@secnumdepth >\m@ne
		\leavevmode \hrule height 0pt depth 0pt
		\large \bfseries \arabic{chapter}\quad % Здесь выводим номер главы, затем пробел. Слово "Chapter" убрано.
		\fi
		\interlinepenalty\@M
		\large \bfseries #1\par}%
	\nobreak
	\vspace*{0pt}%
}
\makeatother

\begin{document}

	\begin{center}
		{\Huge\bf Fusion of light nuclei in a multicluster realization of the three-body problem } 
	\end{center}
	\vspace{0.3cm}

	\centerline{\large\bf Egorov Mikhail~Viktorovich$^{\dag}$.\footnote{egorovphys@mail.ru, https://orcid.org/0000-0001-8231-3884}}
	{\begin{center}
		 Physics Faculty, Tomsk State University, Tomsk, Russia \\ 	
	\end{center}}
	
	\vspace{0.1cm}
	
	\begin{center}
		\begin{minipage}{\textwidth}
			\hspace{0.5cm}
		This work describes a few-body dynamics method based on the Faddeev integral equations in momentum space for determining the total cross sections of fusion and breakup reactions with two- and three-body final channels in the continuum, employing a cluster representation of the colliding nuclei. 
		Total cross sections were obtained for the reactions 
		$^3\text{He}(T,D)^4\text{He}$, $^3\text{He}(T,np)^4\text{He}$, $^3\text{He}(T,nD)^3\text{He}$,
		$^3\text{He}(^3\text{He},2p)^4\text{He}$,   $^3\text{He}(^3\text{He},pD)^3\text{He}$,
		$^7\text{Li}(^3\text{He},\phantom{0}^4\text{He})^6\text{Li}$, 	  $^7\text{Li}(^3\text{He},D^4\text{He})^4\text{He}$, and
		$^7\text{Li}(^3\text{He},T^3\text{He})^4\text{He}$,  in which both the projectile and the target nucleus were treated in a cluster representation. 
		The work also implements a two-potential method to determine the Coulomb $t-$matrix and to account for Coulomb effects in short-range dynamics in momentum space. Calculations of the initial-state Coulomb interaction were performed; furthermore, an estimate was obtained for the magnitude of the off-shell effect of the Coulomb $t-$matrix, as well as the magnitude of the atomic electron anti-screening effect on the Coulomb interaction of the colliding nuclei. 
		The calculated contribution of the cluster mechanism to the total cross section of the considered fusion reactions in the kinetic energy range $T\in[1~\text{keV},20~\text{MeV}]$  is in good agreement with known experimental data.
		\end{minipage}
	\end{center}

	\chapter{Introduction}\label{Introduction}

	Nuclear fusion remains a subject of enduring interest for prospective energy applications \cite{IAEA2024}, for the investigation of nuclei at the neutron drip line \cite{Peni2021}, and for the validation of microscopic reaction models (see reviews \cite{CDCC22}, \cite{SANCSM}  and references therein). For nuclear reactions involving light nuclei at low energies $E\le 1$~MeV, the role of the long-range part of the Coulomb interaction \cite{GloeckleCoul}, \cite{Kouzakov}, \cite{Deltuva} remains a fundamental issue. This includes the effect of atomic electron screening (for instance, in d$^3$He fusion \cite{Aliotta}, \cite{Engstler}, \cite{Pratti}, \cite{Bracci}) and the implementation of few-body dynamics itself, which necessitates the use of rigorous methods from the quantum theory of few-body systems in the energy continuum.
	
	Without delving into the details of the various approaches to describing nuclear reactions with explicit consideration of the few-body aspect, it should be noted that the few-body scattering problem was solved most consistently using the Faddeev equations \cite{Fadd}. These were subsequently generalized to an arbitrary number of bodies \cite{Yakub} and to long-range Coulomb potentials \cite{Merk} in the three-body configuration space.
	
	The merging  of cluster models of nuclei (see an insightful review on the development of cluster models \cite{Lombardo}) with a rigorous treatment of the dynamics of systems containing more than two clusters allows the application of few-body dynamics methods not only to three-nucleon nuclei and nucleon-deuteron scattering but also to more complex nuclei that admit a cluster representation in their ground state. The use of microscopic cluster approaches is primarily presented in the context of studying the internal structure of light and medium nuclei regarding $\alpha$-clustering; the probability of target nucleus breakup into its constituent clusters is evaluated less frequently, and calculations of fusion and rearrangement cross sections in the energy region $E<20-30$~MeV are virtually absent. Optical models with cluster decomposition for both the target nucleus  \cite{Perrotta} and the projectile \cite{GarridoGomez}, based on a systematic analysis of experimental data, highlight the regions where the Distorted Wave Born Approximation (DWBA) fails to adequately capture the specific features of reaction dynamics.

	The description of complex nuclei interactions through the interaction of their individual configurations, starting from single-nucleon ones, has been established since the emergence of the Resonating Group Method (RGM) \cite{Wheeler}, \cite{Wildermuth}. This method initiated the development of microscopic $\alpha$-cluster models of nuclear reactions \cite{Hiura}, including those accounting for Pauli blocking \cite{Saito} and employing generator coordinates \cite{Descovemont}. Within the RGM framework \cite{Fujiwara}, the expansion of the complex nucleus wave function into cluster subgroups enables the description of rearrangement reactions for nuclei with mass numbers up to 10 involving multiple available two-body channels. The description of multichannel reactions, such as those occurring in d + $^7$Li and d + $^7$Be collisions \cite{Lashko}, using the RGM with a cluster representation of the target nucleus, allows by reducing the problem to an effective two-body one to adequately reproduce not only the spectra of the target nuclei but also the astrophysical S-factor of the reactions under consideration. The integration \cite{RGMNCSM} of the RGM with microscopic nuclear models based on $NN$ potentials, including effective field theory, as well as the No-Core Shell Model (NCSM), enables the elevation of cluster-based reaction treatments to a microscopic level. One can also note the successful combination of the Faddeev integral equations, used to determine the cluster wave functions of the $^{11}\text{Li}$ halo nucleus, with the Glauber scattering model at 60~MeV/nucleon, which accurately captures the differential quasielastic scattering of $^{11}\text{Li}$ by $^{12}\text{C}$ nuclei.

	For the study of reactions between light nuclei, the Continuum Discretized Coupled Channels (CDCC) method \cite{CDCC}, \cite{Yahiro} should be highlighted, as it has proven effective in describing cross sections for elastic scattering, breakup, and fusion of weakly bound light nuclei \cite{Souza} while accounting for Coulomb effects. This approach utilizes cluster partitioning for the target and/or the projectile, subsequently reducing the problem to an effective two-body one. For light nuclei, this approach has been successfully applied to neutron- and proton-induced breakup reactions of lithium-7 and lithium-6 isotopes \cite{Guo}, \cite{Ichinkhorlo} into $\alpha-t$ and $\alpha-d$ cluster configurations, which describe the ground states of these nuclei, respectively. The complexity in describing fusion and rearrangement reactions is dictated by the selection of appropriate boundary conditions in configuration space for the eigenfunctions of the channel Hamiltonians. The problem of choosing boundary conditions in the CDCC method can be somewhat compensated for by introducing an imaginary part into the potentials \cite{DiezTorez}. However, the applicability of the CDCC method for a unified description of rearrangement and breakup reactions into three or more fragments remains an open question.
	
	The technique of expanding the multicomponent wave function over a complete set of Gaussian functions \cite{Hiyama}, \cite{Ogawa}, which are eigenfunctions of all channel Hamiltonians of the interacting nuclei, allows for minimizing the issue of matching asymptotic conditions for various reaction channels by increasing the number of coupled channels and explicitly utilizing all equivalent sets of Jacobi variables. This enables the treatment of two-body and many-body channels in a unified manner. The procedure for determining the matrix elements of the transition operator is analogous to solving an eigenvalue problem with the variation of unknown complex constants, making this Gaussian basis expansion method unified and applicable to various problems, yet highly sensitive to the basis size. This approach has been extensively developed for describing the properties of light nuclei, including neutron-rich nuclei, and for determining the binding energies of hypernuclei \cite{Hiyama97}, \cite{Hiyama09} using the cluster representation.
	
	In momentum space, the Faddeev integral equations for the three-body scattering matrix do not explicitly depend on the boundary conditions in the cluster channels and are determined solely by the two-body scattering $t$-matrices. These are constructed in accordance with the Pauli principle and correctly account for the coupling between two-body channels, as well as the energy shift in the free resolvents $G_0$. Such calculations have been previously performed for neutron- and deuteron-induced reactions on lithium nuclei \cite{EgorovnLi}, \cite{EgorovdLi}, $^{12}\text{C}$, and $^{16}\text{O}$ \cite{Deltuva09}.
	
	This work continues the investigation of the cluster structure of light nuclei in fusion reactions by solving a multicluster system of Faddeev integral equations in three-dimensional vector form \cite{EgorovFBSY}. Particular attention is given to the treatment of Coulomb repulsion in the energy region $E < 1$~MeV and to evaluating the influence of the anti-screening effect by a single atomic electron. The study examines the fusion reactions $T(D,n)^4\text{He}$ and $^3\text{He}(D,p)^4\text{He}$, where the target nucleus is represented as a bound $n-d$ or $p-d$ system, respectively, as well as more complex reactions $^3\text{He}(T,np)^4\text{He}$, $^3\text{He}(^3\text{He},2p)^4\text{He}$, $^7\text{Li}(^3\text{He},^4\text{He})^6\text{Li}$, and $^7\text{Li}(^3\text{He},D^4\text{He})^4\text{He}$, in which both the target and the projectile are treated using cluster representations. Total cross sections calculated within a cluster modeling framework based on coupled multicluster Faddeev equations providing a unified description of elastic scattering, rearrangement, and three-body breakup are compared with experimental data. The paper is structured as follows: the second chapter introduces the fundamental elements of three-body dynamics in momentum space and outlines the models and techniques for obtaining $t$-matrices in the two-body sectors; the third chapter provides assessments of the Coulomb interaction contribution to the total cross section and the role of screening parameters. Finally, the resulting cross sections for two- and three-body channels in the energy continuum, accounting for the coupling between various cluster channels and Coulomb effects, are presented in the  fourth chapter.

	\chapter{Dynamical equations for the cluster representation}\label{chapter1}

	In this work, a three-body model is developed for the reactions $^3\text{He}(T,np)^4\text{He}$, $^3\text{He}(^3\text{He},2p)^4\text{He}$, $^7\text{Li}(^3\text{He},^4\text{He})^6\text{Li}$, and $^7\text{Li}(^3\text{He},D^4\text{He})^4\text{He}$ in a multicluster representation, the primary building blocks of which are the fusion processes $T(D,n)^4\text{He}$ and $^3\text{He}(D,p)^4\text{He}$. These latter two processes are treated at a lower two-body level as a basis for the transition to three-body dynamics, where the $^3\text{He}$ nucleus is considered a bound $pD$ system and the triton $T$ as an $nD$ system, respectively, in the reactions $^3\text{He}(T,np)^4\text{He}$ and $^3\text{He}(^3\text{He},2p)^4\text{He}$. In the case of $^3\text{He} + ^3\text{He}$ fusion, the contribution to the cross section arises solely from the $^3\text{He}(D,p)^4\text{He}$ channel, and this contribution is doubled; in contrast, for $^3\text{He} + T$ fusion, both the $T(D,n)^4\text{He}$ and $^3\text{He}(D,p)^4\text{He}$ reaction channels are accessible. It is noteworthy that in the three-body treatment of fusion reactions with a cluster representation of the colliding nuclei, the channels $^3\text{He}(T,D)^4\text{He}$ and $^7\text{Li}(^3\text{He},^4\text{He})^6\text{Li}$ are rearrangement channels belonging to the same system of equations that describes the elastic channel and the three-body breakup channel. Therefore, examining these channels is of interest both for validating the model's performance and for obtaining theoretical total cross sections for the $^7\text{Li}(^3\text{He},^4\text{He})^6\text{Li}$ process, for which experimental data in the energy range $E < 20$~MeV are very scarce \cite{Forsyth}.

	\section{Process kinematics}
	
	When investigating processes of the form 
	\begin{equation}\label{eq1}
		1(E_1,{\bf q}_0)+(23)(W-E_1,-{\bf q}_0)\to 1^{'}(E_1^{'},{\bf q})+2^{'}(E_2^{'},{\bf p}_2^{'})+3^{'}(W-E_1^{'}-E_2^{'},{\bf p}_3^{'}) 
	\end{equation}
	at a fixed total energy $W$ and initial relative momentum ${\bf q}_0$, it is crucial to account for the fact that the (23) subsystem is in a bound (nuclear or atomic) state, and the emission of particles $1'$, $2'$, and $3'$ is accompanied by an energy release $Q > 0$ for the thermonuclear reactions under consideration. The energies and momenta of the particles in (\ref{eq1}) are given in parentheses. Thus, for the mass $M_t$ of the (23) system and the kinetic energy $T$ of the projectile (in the laboratory frame), the invariant mass of the $2'3'$ subsystem varies in the range 
	\begin{equation}\label{eq2}
		\omega_{23}^{'}\in \Big[m_2^{'}+m_3^{'}+Q, m_1+M_t + T\cdot \frac{M_t}{M_t+m_1} - m_1^{'}\Big].
	\end{equation}    
	Specifying the invariant masses (\ref{eq2}) allows for the correct determination of both the energy $E_1^{'}$ of particle $1'$ and the relative momentum $p_{23}^{'}$ of any particle in the $2'3'$ subsystem:
	\begin{align}\label{eq3}
		\begin{gathered}
			E_1^{'}=\frac{W^2+m_1^{'2}-\omega_{23}^{'2}}{2W}, \\
			p_{23}^{'}=\frac{\sqrt{(\omega_{23}^{'2}-(m_2^{'}-m_3^{'})^2)(\omega_{23}^{'2}-(m_2^{'}+m_3^{'})^2)}}{2\omega_{23}^{'}}.  
		\end{gathered}
	\end{align} 
	The quantities in (\ref{eq3}) are necessary for performing the Lorentz transformation from the $2'3'$ subsystem to the overall center-of-mass frame. The momenta ${\bf p}_2^{'}$ and ${\bf p}_3^{'}$ of particles $2^{'}$ and $3^{'}$ in the overall center-of-mass frame are obtained using Lorentz transformations:
	\begin{equation}\label{eq4}
		{\bf p}_{2'3'} = \pm {\bf p}_{23}^{'} +
		\Big(-\frac{{\bf q}}{\omega_{23}^{'}}\Big)\Big[\frac{\mp {\bf p}_{23}^{'}{\bf q}}{W-E_1^{'}+\omega_{23}^{'}}+\frac{\omega_{23}^{'}}{2}\Big]
	\end{equation}
	In (\ref{eq4}), the upper and lower signs correspond to particles $2^{'}$ and $3^{'}$, respectively. The correct reconstruction of the momenta of all final particles emitted into the energy continuum is necessary to determine the Lorentz-invariant cross section of process (\ref{eq1}):
	\begin{equation}\label{eq5}
		\frac{d^3\sigma}{d\omega_{23}^{'}d\Omega_{23}^{'}d\Omega_1} = 
		\frac{E_1(W-E_1)q p_{23}^{'} E_1^{'}E_2^{'}E_3^{'}}{q_0 W^2 (2\pi)^5}
		\frac{\sum\big|U_0({\bf p},{\bf q};{\bf q}_0)\big|^2}{(2j_1+1)(2j_2+1)}.
	\end{equation}
	The summation of the squared breakup amplitude $U_0({\bf p},{\bf q};{\bf q}_0)$ is performed over the spin projections of the final particles. For potentials that do not contain operators acting in the angular momentum space, the summation in (\ref{eq5}) can be carried out explicitly. The spins of the colliding nuclei in (\ref{eq5}) are denoted by $j_1$ and $j_2$.
	
	For the simpler case of elastic scattering or a rearrangement reaction of the form
	\begin{equation}\label{eq6}
		1(E_1,{\bf q}_0)+2(W-E_1,-{\bf q}_0) \to 
		1^{'}(E_1^{'},{\bf q}) + 2^{'}(W-E_1^{'},-{\bf q})
	\end{equation}
	at a total energy $W$, the total cross section is determined by integrating the formula
	\begin{equation}\label{eq7}
		\frac{d\sigma}{d\Omega}=
		\frac{E_1(W-E_1)E_1^{'}(W-E_1^{'})}{(2\pi W)^2}\frac{q}{q_0}
		\frac{\sum \big|U({\bf q},{\bf q}_0)\big|^2}{(2j_1+1)(2j_2+1)}
	\end{equation}
	over the solid angle of the outgoing particles $1^{'}$ and $2^{'}$. The amplitude $U({\bf q},{\bf q}_0)$ for process (\ref{eq6}) is summed over the spin projections of the final particles. The determination of the amplitudes $U_0({\bf p},{\bf q};{\bf q}_0)$ and $U({\bf q},{\bf q}_0)$ is addressed in the following sections.

	\section{Faddeev equations for thermonuclear reactions}
	
	The interaction of an incident $^3\text{He}$ nucleus with a triton or a $^3\text{He}$ nucleus acting as a target which, in the cluster representation, is a bound neutron-deuteron (or proton-deuteron) system requires the determination of several intermediate scattering $t$-matrices. The following scheme, using the $^3\text{He}(T,np)^4\text{He}$ reaction as an example, illustrates how the partition index $i\in [1,2,3]$ (identifying the spectator particle) and the indices $\alpha, \beta$ for the channel representations are assigned to specific nuclei:
	\begin{equation}\label{eq8}
		\begin{matrix}
			\phantom{0} \\
			\alpha \\
			\beta
		\end{matrix}
		\begin{pmatrix}
			1 & 2 & 3 \\
			\phantom{0}^3\text{He} & n & D \\
			p & n & \phantom{0}^4\text{He}
		\end{pmatrix}:
		\,\, \phantom{0}^3\text{He} + (nD) \to n + p + \phantom{0}^4\text{He}
	\end{equation} 
	A similar scheme for the cluster representation of the $^3\text{He}(^3\text{He},2p)^4\text{He}$ reaction is obtained by replacing the neutron with a proton in channels $\alpha$ and $\beta$.
	
	A distinctive feature of these calculations is that, even at the stage of preparing the two-body scattering $t$-matrices corresponding to scheme (\ref{eq8}), a three-body system of Faddeev equations was used to determine the elastic $nD$ scattering amplitudes with the Bonn NN potential \cite{BonnNN}. Calculations performed in \cite{EgorovFBSY} showed that a [2/2] Pad\'e approximant is sufficient to describe elastic $nD$ scattering. The resulting amplitudes $U({\bf q}_0,{\bf q})$ were subsequently stored as numerical tables containing the amplitude values, the magnitudes of the relative momenta before (${\bf q}_0$) and after (${\bf q}$) scattering, and the polar and azimuthal angles between them. No distinction was made between $pD$ and $nD$ interactions at this stage. Consequently, for the calculations of the $^3\text{He}(T,np)^4\text{He}$ and $^3\text{He}(^3\text{He},2p)^4\text{He}$ fusion reactions, the elastic $nD$ scattering amplitudes $U({\bf q}_0,{\bf q})$ were retrieved from arrays previously stored in the computing environment's shared memory, using a nearest-point selection criterion on the numerical grid of momenta ${\bf q}_0$ and ${\bf q}$.
	
	To determine the cross section (\ref{eq5}), it is necessary to link the breakup amplitude $U_0({\bf p},{\bf q};{\bf q}_0)$ to the three-body scattering $T$-matrices. Since the partition index $i$ of the breakup $T$-matrices is associated with the particle exiting the system, we can directly write in the momentum representation:
	\begin{equation}\label{eq9}
		U_0({\bf p},{\bf q}; q_0) =
		\big\langle {\bf p} {\bf q}\big|T_1\Psi_{1(23)}\big\rangle 
		+  
		\big\langle \tilde{\bf p} \tilde{\bf q}\big|T_2\Psi_{2(31)}\big\rangle
		+
		\big\langle {\bar{\bf p}} {\bar {\bf q}}\big|T_3\Psi_{3(12)}\big\rangle
	\end{equation}
	The momenta ${\bf p}$ and ${\bf q}$ in (\ref{eq9}) are associated with the corresponding sets of Jacobi variables and represent the relative momentum within a chosen pair of particles and the momentum of the third particle relative to that pair. It was shown in \cite{EgorovFBSY} that the momentum pairs $(\tilde{\bf p}, \tilde{\bf q})$ and $(\bar{\bf p}, \bar{\bf q})$ required for calculating the breakup matrices $T_2$ and $T_3$ can be obtained by applying particle permutation operators to the expressions for ${\bf p}$ and ${\bf q}$. Direct comparison confirms that these same momenta can be derived from the overall center-of-mass system by defining all three sets of Jacobi variables. The eigenfunctions $\Psi_{i(jk)}$ in (\ref{eq9}) define the specific subsystem $(jk)$ in the initial state, which may include the eigenfunction of the $(jk)$ subsystem with a specific binding energy if the subsystem admits such bound states. In general, the functions $\Psi_{i(jk)}$ can be treated not only as plane-wave asymptotic states of the spectator particle and the $(jk)$ subsystem but also as moving wave packets.

	From the standpoint of three-body dynamics and the general unitarity of the three-body scattering matrix, processes involving two and three particles in the final state are interconnected. This implies that the matrix elements of the rearrangement operator $U_{ji}$, which describe transitions from two-body states to other two-body states and satisfy their own system of coupled integral equations, can be obtained \cite{Gloeckle3N} from the Faddeev breakup $T$-matrices using simple quadrature formulas:
	\begin{equation}\label{eq10}
		\langle \Psi_{j(ki)}| U_{ji} |\Psi_{i(jk)}\rangle =
		\langle\Psi_{j(ki)}| P G_0^{-1} + P T_{1}|\Psi_{i(jk)}\rangle.
	\end{equation} 
	In three-dimensional vector notation, equation (\ref{eq10}) for elastic scattering takes the form:
	\begin{align}\label{eq11}
		\begin{gathered}
			\langle \Psi_{1(23)}| U_{11}| \Psi_{1(23)}\rangle =
			\phi_{(23)}\big(-{\bf q}\tfrac{m_1}{m_1+m_3}-{\bf q}_0\big)\big(E-\tfrac{q^2}{2\mu_{23}}-\tfrac{q_0^2}{2\mu_{13}}-
			\tfrac{q_0q y_{qq_0}}{m_3} \big)\phi_{(31)}\big({\bf q}+{\bf q}_0\tfrac{m_2}{m_2+m_3}\big)  \\
			+
			\phi_{(23)}\big({\bf q}\tfrac{m_1}{m_1+m_2}+{\bf q}_0\big)\big(E-\tfrac{q^2}{2\mu_{23}}-\tfrac{q_0^2}{2\mu_{12}}-
			\tfrac{q_0q y_{qq_0}}{m_2} \big)\phi_{(12)}\big(-{\bf q}-{\bf q}_0\tfrac{m_3}{m_2+m_3}\big)+\\
			\int d^3 q^{'}\Big[
			\langle\Psi_{1(23)}| -{\bf q}\tfrac{m_1}{m_1+m_3}-{\bf q}^{'}, {\bf q}^{'}\rangle\langle {\bf q}+{\bf q}^{'}\tfrac{m_2}{m_2+m_3}, {\bf q}^{'}|T_2|\Psi_{2(31)}\rangle+ \\
			\langle\Psi_{1(23)}| {\bf q}\tfrac{m_1}{m_1+m_2}+{\bf q}^{'}, {\bf q}^{'}\rangle\langle -{\bf q}-{\bf q}^{'}\tfrac{m_3}{m_2+m_3}, {\bf q}^{'}|T_3|\Psi_{3(12)}\rangle
			\Big].
		\end{gathered}
	\end{align}
	The eigenfunctions of the two-body subsystems are denoted as $\phi$ in (\ref{eq11}). By the action of the permutation operators $P$ (\ref{eq10}) of the partition indices on the inverse resolvents $G_0^{-1}$, we obtain two terms corresponding to transitions between paired bound states: $\phi_{(31)}\to \phi_{(23)}$ and $\phi_{(12)}\to \phi_{(23)}$. The transition to rearrangement processes is accompanied not only by a change in the momentum magnitude $q \neq q_0$ of the scattered particle but also by the substitution of the functions $\Psi_{1(23)}\to\Psi_{2(31)}$ and $\Psi_{1(23)}\to\Psi_{3(12)}$, depending on which particle index $j=2$ or $j=3$ is to be fixed in the final state. For the incident particle $i=1$, the momentum dependence of all functions in (\ref{eq11}) and the particle mass indices $m_1, m_2, m_3$ remain identical to those in the elastic process, which advantageously distinguishes the use of Faddeev equations for describing reactions.
	
	The transition to channel representations involving different particle types, according to scheme (\ref{eq8}), follows the rule $t_i\to t_i^{\beta\alpha}$, $U_{ji}\to U_{ji}^{\beta\alpha}$, and $T_i\to T_i^{\beta\alpha}$, due to the fact that the two-body $t_i$-matrices also become matrices in the space of two-body channels. From the viewpoint of thermonuclear reactions, the most important channel is $i=2$, in which the thermonuclear reaction itself occurs, for example, $^3\text{He}(D,p)^{4}\text{He}$. Furthermore, the Faddeev equation system and the Faddeev approach, in general, are equally well-adapted for incorporating all possible rearrangement processes and reactions in each of the two-body channels, resulting in an increased number of coupled equations. Assuming for simplicity only one reaction for the partition index $i=2$, the system of Faddeev integral equations takes the form:
	\begin{equation}\label{eq12}
		\begin{pmatrix}
			T_1^{11} \\
			T_1^{21} \\
			T_2^{11} \\
			T_2^{21} \\
			T_3^{11} \\
			T_3^{21} 
		\end{pmatrix}=
		\begin{pmatrix}
			t_1^{11}(\phi_2+\phi_3) \\
			t_1^{21}(\phi_2+\phi_3) \\
			t_2^{11}(\phi_3+\phi_1)  \\
			t_2^{21}(\phi_3+\phi_1)  \\
			t_3^{11}(\phi_1+\phi_2)  \\
			t_3^{21}(\phi_1+\phi_2)   
		\end{pmatrix}
		+
		\begin{pmatrix}
			0 & 0 &  t_1^{11}G_0 & 0 & t_1^{11}G_0 & 0 \\
			0 & 0 & 0            & t_1^{22}G_0 & 0 & t_1^{22}G_0 \\
			t_2^{11}G_0 & t_2^{12}G_0 & 0 & 0 &  t_2^{11}G_0 & t_2^{12}G_0 \\
			t_2^{21}G_0 & t_2^{22}  & 0 & 0 & t_2^{21}G_0 & t_2^{22}G_0 \\
			t_3^{11}G_0 &  0      & t_3^{11}G_0 & 0 & 0 & 0 \\
			0 & t_3^{22}G_0 & 0 & t_3^{22}G_0  & 0 & 0  
		\end{pmatrix}
		\begin{pmatrix}
			T_1^{11} \\
			T_1^{21} \\
			T_2^{11} \\
			T_2^{21} \\
			T_3^{11} \\
			T_3^{21} 
		\end{pmatrix}.
	\end{equation}
	Identical superscripts in (\ref{eq12}) correspond to processes involving the emission of particles without a change in the particle species, whereas a change in the representation space $1\to 2$ corresponds to a change in particle types, i.e., to thermonuclear reactions within the cluster representation of nuclei. Thus, it is evident that for the reactions under consideration, the quantities $T_i^{21}$ with $i\in[1,2,3]$ are of primary interest. The system (\ref{eq12}) is symmetric with respect to the permutation of particle indices; for the specific process (\ref{eq8}), possessing only a single bound $nD$ state, we have $\phi_2=\phi_3=0$. The eigenfunctions of the two-body subsystems $\phi_i\equiv\phi_{jk}$ for $i\not=j\not=k$ in (\ref{eq12}) are labeled according to the spectator particle indices.
	
	The two-body Green's functions $G_0$ in (\ref{eq12}) exhibit an energy shift associated with the motion of the spectator particle for each partition and channel number. This shift qualitatively distinguishes three-body dynamics from the motion of a particle in an effective field. This shift, which varies for each set of Jacobi variables, determines the scattering energy $z$ in the two-body subsystem:
	\begin{align}\label{eq13}
		\begin{gathered}
			z=E-\frac{q_{i(jk)}^2}{2M_{i(jk)}}, \\
			E=T\frac{M_t}{M_t+m_1}-|E_b|.
		\end{gathered}
	\end{align}     
	As seen from formula (\ref{eq13}), the three-body scattering energy $E$ is related to the kinetic energy of the incident particle and reflects the dependence on the binding energy $E_b$ of the initial subsystem $(23)$ in processes (\ref{eq8}). Expression (\ref{eq13}) also reveals that the quantities $E$ and $z$ can be either positive or negative, which alters the distribution of two-body poles on the complex energy plane during the transition to the three-body phase space. The presence of a third particle in the interaction region modifies the branch-cut region of the resolvents, deforming them into so called moon like regions, known as regions of logarithmic singularities. By writing any of the resolvents $G_0$ in the form $G_0\equiv m/(qq^{''})/(y_0-x^{''})$, where $q^{''}$ is the radial integration variable and $x^{''}$ is the cosine of the polar angle in the integral of (\ref{eq12}), an expression explicitly containing a logarithmic singularity can be obtained for any iteration of equation (\ref{eq12}):
	\begin{align}\label{eq14}
		\begin{gathered}
			\int q^{''2}dq^{''} d\Omega_{q^{''}} t \frac{m}{qq^{''}}\frac{1}{y_0-x^{''}}t
			\equiv \int q^{''} dq^{''} \int\limits_{-1}^{1}
			dx^{''}\frac{f(q^{''},x^{''})}{y_0-x+i\varepsilon} =  \\
			\int q^{''} dq^{''} \Big[  
			\int\limits_{-1}^{1} dx^{''}\frac{f(q^{''},x^{''})-\hat f(q^{''},y_0)}{y_0-x^{''}} 
			+\int\limits_{-1}^{1} dx^{''}\frac{\hat f(q^{''},y_0)}{y_0-x^{''}}
			\Big] \equiv \\
			\int q^{''}dq^{''}\Big[ 
			\int\limits_{-1}^{1} dx^{''}\frac{f(q^{''},x^{''})-\hat f(q^{''},y_0)}{y_0-x^{''}}
			+\hat f(q^{''},y_0) \ln{\Big(\Big|\frac{1+y_0}{1-y_0}\Big|\Big)} -\Theta(1-|y_0|)i\pi \hat f(q^{''},y_0)
			\Big].
		\end{gathered}
	\end{align}
	The arguments and indices of the two-body $t$-matrices on the left hand side of (\ref{eq14}) are omitted. The quantity $\hat f(q^{''},y_0)$ is equal to the original function $f(q^{''},y_0)$ when $|y_0|\le 1$, and equals $f(q^{''},y_0/|y_0|)$ when $|y_0|>1$. The Heaviside step function $\Theta$ in (\ref{eq14}) governs the entry into the domain $|y_0|\le 1$, where a pole contribution appears in the integral (\ref{eq14}). It should be noted that the final two terms in the last row of (\ref{eq14}) arise in non overlapping kinematic situations. Handling the regions of logarithmic singularities during integration is more conveniently performed using spline interpolation while monitoring the boundaries of these regions.
	
	The presence of rearrangement channels in the three-body system implies that in the region $E > 0$, the algebraic approximation of the homogeneous form of the integral equation (\ref{eq12}) admits non trivial solutions (i.e., bound states). Consequently, when transforming equation (\ref{eq12}) into a matrix form on a numerical momentum grid, the system cannot be solved by direct matrix inversion. Therefore, the solution to the system (\ref{eq12}) is sought in the form of successive iterations:
	\begin{equation}\label{eq15}
		T_i^{\beta\alpha}\approx \sum\limits_n\big(C_n\big)_i^{\beta\alpha}
	\end{equation}
	where $n=0$ corresponds to the inhomogeneous term of the system (\ref{eq12}). Each iteration in (\ref{eq15}) satisfies a system of coupled recurrence relations:
	\begin{align}\label{eq16}
		\begin{gathered} 
			\big(C_n\big)_1^{11} = t_1^{11} G_0\big(C_{n-1}\big)_2^{11} +t_1^{11}G_0\big(C_{n-1}\big)_3^{11}\\
			\big(C_n\big)_1^{21} = t_1^{22} G_0 \big(C_{n-1}\big)_2^{21}
			+t_1^{22}G_0 \big(C_{n-1}\big)_3^{21} \\
			\big(C_n\big)_2^{11} = t_2^{11}G_0\Big[\big(C_{n-1}\big)_1^{11}+\big(C_{n-1}\big)_3^{11}\Big]
			+t_2^{12}G_0 \Big[\big(C_{n-1}\big)_1^{21}+\big(C_{n-1}\big)_3^{21}\Big] \\
			\big(C_n\big)_2^{21} = t_2^{21}G_0\Big[\big(C_{n-1}\big)_1^{11}+\big(C_{n-1}\big)_3^{11}\Big]
			+t_2^{22}G_0 \Big[\big(C_{n-1}\big)_1^{21}+\big(C_{n-1}\big)_3^{21}\Big] \\
			\big(C_n\big)_3^{11} = t_3^{11} G_0\big(C_{n-1}\big)_1^{11} +t_3^{11}G_0\big(C_{n-1}\big)_2^{11} \\
			\big(C_n\big)_3^{21} = t_3^{22} G_0\big(C_{n-1}\big)_1^{21} +t_3^{21}G_0\big(C_{n-1}\big)_2^{21}. 
		\end{gathered}
	\end{align}
	It is evident from the system (\ref{eq16}) that the coupling between channels involving different particle species manifests only starting from $n=2$. In the present work, the direct solution of (\ref{eq16}) was performed up to $n=2$ inclusive.

	\section{Two-body $t$-matrices}
	
	Two-body scattering $t$-matrices are the fundamental building blocks for investigating fusion and scattering reactions at both the two-body and three-body problem levels. Moreover, the use of $t$-matrices in few-body dynamics obviates the need for selecting appropriate boundary conditions for wave functions in fusion and rearrangement reactions, a problem that arises in the coordinate representation.  
	For the coupled fusion and elastic scattering channels, this work utilizes local potentials of the form
	\begin{equation}\label{eq17}
		V({\bf p}_f,{\bf p}_i) =  \frac{V_R}{\mu_R^2+p_f^2+p_i^2-2p_fp_i y_{p_fp_i}}  
	\end{equation}  
	with two free parameters, $V_R$ and $\mu_R$. The cosine of the angle between the vectors ${\bf p}_f$ (defined by polar $\theta_f$ and azimuthal $\phi_f$ angles) and ${\bf p_i}$ (defined by polar $\theta_i$ and azimuthal $\phi_i$ angles) is given by
	\begin{equation}\label{eq18}
		y_{p_fp_i}=\cos{\big(\theta_i\big)}\cos{\big(\theta_f\big)}
		+\sqrt{1-\cos^2{\big(\theta_i\big)}}\sqrt{1-\cos^2{\big(\theta_f\big)}}
		\cos{\big(\phi_f+\phi_i\big)}.  
	\end{equation} 
	Taking the $D^3$He collision as an example, we denote the interactions $V_{ij}$ in channels $i,j=1,2$ as follows:
	\begin{equation}\label{eq19}
		\begin{cases}
			V_{11}:\,\phantom{0}^3\text{He}(D,D)^3\text{He} \\
			V_{12}:\,\phantom{0}^3\text{He}(D,p)^4\text{He} \\
			V_{21}:\,\phantom{0}^4\text{He}(p,D)^3\text{He} \\
			V_{22}:\,\phantom{0}^4\text{He}(p,p)^4\text{He}. \\
		\end{cases}
	\end{equation}
	The potentials (\ref{eq17}) are used to describe the elastic scattering ($V_{11}$) of the colliding nuclei, as well as the forward and backward reactions ($V_{21}$, $V_{12}$) between them. The dynamical Lippmann-Schwinger equations for the scattering matrices $t_{11}$ and $t_{21}$, which are of primary interest in this work, take the form:
	\begin{equation}\label{eq20}
		\begin{cases}
			t_{11}=V_{11}+V_{11}G_0t_{11}+V_{12}G_0t_{21} \\
			t_{21}=V_{21}+V_{21}G_0t_{11}+V_{22}G_0t_{21}. \\
		\end{cases}  
	\end{equation}
	It is evident from the system (\ref{eq20}) that the potentials $V_{22}$, given unknown $V_{21}$, can be parameterized independently based on well-established partial phase shifts for elastic $n^3$He and $n^4$He scattering. For the potentials $V_{22}$ acting in the partial wave $L$ with total angular momentum $J$, a separable form is employed:
	\begin{align}\label{eq21}
		\begin{gathered}
			V_L^J(p_f,p_i)=\lambda_L^J\xi_L(p_f)\xi_L(p_i), \\
			\xi_L(p)=\frac{c_1p^L}{(p^2+b_1^2)^{L+1}}+\frac{c_2p^{L+2}}{(p^2+b_2^2)^{L+2}},  
		\end{gathered}
	\end{align}
	where the unknown constants $c_1, c_2, b_1, b_2$ and interaction strengths $\lambda_L^J$ are determined by fitting to known phase shifts for the lowest partial waves. The parameters for the local potentials (\ref{eq17}) were selected to reproduce the total cross sections of the considered fusion reactions in the energy range $T < 20$~MeV. The resulting potential parameters are presented in Table~(\ref{tab1}). It should be noted that the separable potentials (\ref{eq21}) generate partial two-body $t$-matrices $T_L^J$, which, in the absence of spin and tensor interactions, are related to the full $t$-matrix via the partial-wave expansion:
	\begin{align}\label{eq22}
		\begin{gathered}
			T({\bf p}_f,{\bf p}_i)=4\pi\sum\limits_{JMLM_iM_f}C_{LM_f\,S_fM-M_f}^{JM} Y_{LM_f}(\hat p_f)\chi_{S_fM-M_f} \\
			T_L^J(p_f,p_i) C_{LM_iS_iM-M_i}^{JM}Y_{LM_i}^*(\hat p_i)\chi_{S_iM-M_i}^*.
		\end{gathered}
	\end{align}
	The summation in (\ref{eq22}) is carried out over the total angular momentum $J$ and its projection $M$, as well as over the orbital angular momentum $L$ and its projections in the initial $M_i$ and final $M_f$ states. The coupling of the spins in the initial $S_i$ and final $S_f$ states with the orbital waves is governed by Clebsch-Gordan coefficients. The spin functions $\chi$ and spherical harmonics $Y(\hat p)$, where $\hat p$ denotes the angular variables of the vector ${\bf p}$, enter (\ref{eq22}) as functions of their respective variables. It should be noted that the parameters of the short-range interactions $V_{ij}$ were determined primarily based on the description of experimental data in the region of total cross section maxima, with the objective of further investigating the relative contribution of the Coulomb interaction, which manifests not only as a separate term in the scattering amplitude but also as a perturbation in the short-range resolvent.
	%%%%%%%%%%%%%%%%%%%%%%%%%%%%%%%%%%%%%%%TABLE1%%%%%%%%%%
	\begin{table}[p]
		\caption{Parameters of short-range separable potentials for elastic $n^4\text{He}$ ($^1S_{1/2}$, $^2P_{1/2}$, $^2P_{3/2}$, $^2D_{3/2}$), $n^3\text{He}$ ($^1S_0$, $^3S_1$, $^1P_1$, $^3P_0$, $^3P_1$, $^3P_2$, $^1D_2$), $T^4\text{He}$ ($^2P_{3/2}$, $^2P_{1/2}$, $^3F_{7/2}$), and $^4\text{He}^4\text{He}$ ($0^+$, $2^+$) interactions in the lowest partial waves (top), and parameters of local potentials for the considered reactions and scattering processes (bottom).}
		\label{tab1}	
		\bigskip
		\begin{tabular}{|c|c|c|c|c|c|}
		%\begin{tabular*}{\tblwidth}{@{}LLLLLL@{}}
			\hline
			$^{2S+1}L_J$ & $c_1,[\text{MeV}^{L+3/2}]$ & $b_1,[\text{MeV}]$ & 
			$c_2,[\text{MeV}^{L+3/2}]$ & $b_2,[\text{MeV}]$  & $\lambda^{-1},[\text{MeV}]$  \\
			\hline
			$^1S_{1/2}$ &  14.1$\cdot 10^3$ & 150.1 & 3060 & 124.1 & W-4668.7 \\
			$^2P_{1/2}$ & 5.88$\cdot 10^5$ & 125.1 & 790 & 1.3 & -10$^{-4}$ \\
			$^2P_{3/2}$ & 85 & 106.1  & 4020 & 200 & -10$^{-4}$ \\
			$^2D_{3/2}$ & 110 & 48 & 2.34$\cdot 10^5$ & 173 & -10$^{-4}$ \\
			\hline
			$^1S_0$ &  40 & 16.1 & 315.1 & 34.1 & 5 \\
			$^3S_1$ & 47.1 & 184 &   31.1 & 57.1 & 2$\cdot 10^{-2}$ \\
			$^1P_1$ & $10^3$ & 1.1 & 4.36$\cdot 10^5$ & 67.1 & 1$\cdot 10^{-2}$ \\
			$^3P_0$ & 18 & 1.1 & 7.5$\cdot 10^3$ & 72.1 & 6$\cdot 10^{-2}$ \\
			$^3P_1$ & 46 & 2.3 & 8010 & 68 & 7$\cdot 10^{-2}$ \\
			$^3P_2$ & 43 & 2.3 & 7830 & 70 & 7$\cdot 10^{-2}$ \\
			$^1D_2$ & 2.1 & 1.1 & 1.12$\cdot 10^6$ & 118 & 2$\cdot 10^{-2}$ \\
			\hline
			$^1S_0$  & $10^5$  & 520 & -3$\cdot 10^4$   &  105 & -500 \\
			$^1D_2$  & 43  & 14  & 15.05$\cdot 10^4$  & 254  & $10^{-7}$ \\
			\hline
			$^2P_{3/2}$ & 1 & 2$\cdot 10^3$ &  246 & 4886 & $10^{-7}$ \\
			$^2P_{1/2}$ & 1 & 2$\cdot 10^3$ &  246 & 4886 & $10^{-7}$ \\
			$^3F_{7/2}$ & 0.2 & 31.1 &  2250 & 223.1 & $10^{-7}$ \\
			\hline
			\multicolumn{4}{|c|}{processes}  & $V_R$  & $\mu_R$ \\
			\hline
			\multicolumn{4}{|c|}{$^3\text{He}(D,D)^3\text{He}$}  &  0.6 & 2.6 \\
			\multicolumn{4}{|c|}{$^3\text{He}(D,p)^4\text{He}$}
			&  3.0 & 8.5 \\ 
			\multicolumn{4}{|c|}{$T(D,D)T$}
			&  6$\cdot 10^{-3}$ & 1.15 \\   
			\multicolumn{4}{|c|}{$T(D,n)^4\text{He}$}
			&  0.31 & 1.51 \\  
			\multicolumn{4}{|c|}{$^7\text{Li}(^3\text{He},\phantom{0}^3\text{He})^7\text{Li}$}
			&  3.0 & 1.0 \\  
			\multicolumn{4}{|c|}{$^7\text{Li}(^3\text{He},\phantom{0}^4\text{He})^6\text{Li}$}
			&  $10^{-2}$ & 0.1 \\ 
			\multicolumn{4}{|c|}{$^6\text{Li}(^4\text{He},\phantom{0}^4\text{He})^6\text{Li}$}
			&  5.0 & 2.0 \\
			\hline
		\end{tabular}
	\end{table}
	%%%%%%%%%%%%%%%%%%%%%%%END TABLE1%%%%%%%%%%%%%%%%%%

	\chapter{Effects of the Coulomb interaction}
	
	In nuclear fusion reactions and nuclear scattering at low and intermediate energies within the considered range $T\in[1~\text{keV}, 20~\text{MeV}]$, the Coulomb interaction manifests in three key aspects:
	\begin{enumerate}
		\item An individual contribution of the Coulomb $t$-matrix $T_C$ to elastic scattering, associated with the repulsion of positively charged nuclei during scattering and the interaction of charged nuclei in the initial (ISI) and final (FSI) states, particularly when the initial or final channels contain at least one neutral particle.
		\item Perturbation of short-range resolvents by the Coulomb $T_C$ scattering matrix, which leads to additional terms in the integral kernels of equations (\ref{eq12},\ref{eq20}). This perturbation reflects the influence of the long-range potential on the dynamics of strongly interacting particles.  
		\item Anti-screening of the reaction by atomic electrons of the target atoms. This effect arises from the impossibility of using bare nuclei without atomic electrons in experiments; it manifests in the energy region $T < 10$~keV and plays a crucial role in the study of astrophysical $S$-factors.  
	\end{enumerate}
	Directly solving the Lippmann-Schwinger equations with a Coulomb potential is known to be hindered by the inability to separate the singularities of the original Coulomb potential from the singularities of the $G_0$ resolvents in the energy continuum. Furthermore, in the coordinate representation, the Coulomb potential does not satisfy the asymptotic condition $\int_0^{\infty}dr\cdot r\cdot |V_C| < \infty$, which prevents a rigorous neglect of the long-range part of the Coulomb interaction. Nevertheless, the solution to the scattering problem in a purely Coulomb potential is well known and is given by
	\begin{equation}\label{eq23}
		\Psi_{\bf k}^C({\bf r}) = e^{i{\bf k}{\bf r}}
		e^{-\pi\eta/2}\Gamma(1+i\eta) F(-i\eta;1;ikr(1-\cos{(\theta)})),
	\end{equation}
	where ${\bf r}$ denotes the coordinate of a particle moving with momentum ${\bf k}$ at an angle $\theta$ relative to the $Oz$ axis, and $\Gamma(1+i\eta)$ is the Gamma function of a complex argument with the Sommerfeld parameter $\eta$. Since the hypergeometric function $F$ in (\ref{eq23}) reduces to unity at $r=0$ or at a zero scattering angle, the squared Coulomb scattering wave function in this case takes the simple form
	\begin{equation}\label{eq24}
		|\Psi_{\bf k}^C({\bf r})|^2_{r\to 0}=\frac{2\pi\eta}{e^{2\pi\eta}-1},
	\end{equation} 
	which, as $\eta\to\infty$, transitions into the Gamow factor, traditionally used to estimate the suppression of fusion cross sections at low energies. 
	
	Solving the Coulomb problem for two- and three-body systems is a non trivial and labor intensive task. It has been implemented for scattering and nuclear reactions using the adiabatic approximation \cite{Adiobat}, the introduction of a tridiagonal basis in the form of Sturm-Coulomb functions \cite{Papp}, and wave-packet continuum discretization techniques \cite{RubtsovaCoul}. This work employs the two-potential method to account for the Coulomb interaction in momentum space, as outlined in points 1–3 above, following the developments in \cite{Oryu06,Oryu16}. Formally, this method is similar to the approach in \cite{Merk}, where the Coulomb potential is partitioned into long-range and short-range parts in configuration space, followed by their investigation using independent Lippmann-Schwinger equations. A comparison of the efficiency of these various methods for describing low-energy nuclear reaction cross sections lies beyond the scope of the present work; therefore, a brief description of the results obtained using the two-potential method \cite{Oryu06} is provided below.
	
	The Coulomb potential $V_C$ in the two-potential method is decomposed into a sum:
	\begin{equation}\label{eq25}
		V_C=V_R+(V_C-V_R)\equiv V_R + V_{\phi},
	\end{equation}
	where the screened part $V_R$ is chosen in the form:
	\begin{equation}\label{eq26}
		V_R({\bf p}_f,{\bf p}_i)=\frac{4\pi k\eta/\mu}{p_f^2-2p_fp_iy_{p_fp_i}+p_i^2+\Big(\frac{\hbar c}{R}\Big)^2}  
	\end{equation}
	with the screening parameter:
	\begin{equation}\label{eq27}
		R=\frac{\hbar c}{2k}\exp{\Big(\frac{C(k,l)}{\eta}\Big)}.
	\end{equation}
	In the general case, the quantity $C(k,l)$ in (\ref{eq27}) can depend on both the orbital angular momentum $l$ and the momentum of the colliding particles $k=\sqrt{2\mu |z|}$, where $\mu$ is the reduced mass of the particles and $z$ is their two-body scattering energy. The choice of the parameter $C(k,l)$ in the formula for the screening radius serves two purposes: i) boundary conditions for the $T_{\phi}$-matrix generated by the potential $V_{\phi}$ in the form
	\begin{equation}\label{eq28}
		T_{\phi}({\bf k},{\bf k})=T_{\phi}({\bf k},{\bf p}_i)=T_{\phi}({\bf p}_f,{\bf k})=0,
	\end{equation}
	and ii) the absence of numerical oscillations when solving the corresponding Lippmann-Schwinger equations for the $T_{\phi}$-matrix. The implementation of the boundary conditions in the form (\ref{eq28}) ensures the suppression of unwanted singularities of the Coulomb potential, which manifest in the region of small relative momenta. As calculations have shown, condition (\ref{eq28}) can be satisfied with an accuracy of $<10^{-10}$ for values of $C(k,l)$ equal to some constant for all partial waves and momenta. In the present calculations, the constant values were chosen as follows: $C=0.5$ for electron-nucleus interactions and $C=-2$ for nucleus-nucleus interactions.
	
	The full Coulomb $T^C$-matrix is obtained as the sum:
	\begin{equation}\label{eq29}
		T_C=T_{\phi}+T_{R\phi},
	\end{equation}
	where the screened $T_{R\phi}$-matrix is defined by the action of M\"oller operators on the $t^{R\phi}$-matrix:
	\begin{equation}\label{eq30}
		T_{R\phi}=(1+T_{\phi}G_0)t^{R\phi}(1+G_0T_{\phi}),
	\end{equation} 
	which in turn is generated by the screened potential $V_R$:
	\begin{equation}\label{eq31}
		t^{R\phi}=V_R+V_RG_{\phi}t^{R\phi}\equiv V_R+K^{sC}G_0t^{R\phi}.
	\end{equation}
	It contains an energy shift in the resolvent $G_{\phi}\equiv G_0+G_0T_{\phi}G_0$, which leads to an additional term in the integral kernel $K^{sC}$: 
	\begin{equation}\label{eq32}
		K^{sC}=V_R+V_RG_0T_{\phi}.
	\end{equation}
	On the mass shell, the identity $T_{R\phi}\equiv t^{R\phi}$ holds; however, for calculating ISI and FSI effects, the Coulomb $T_C$-matrix (\ref{eq30}) is required half off-shell. This implies that, even in the two-body problem, the total number of terms in the $T_C$-matrix reaches five, with multiple integrals contained in four of them, considering (\ref{eq31},\ref{eq32}). In this regard, the use of an off-shell two-potential approach is computationally burdensome, with costs significantly increasing in the sector with three interacting particles \cite{Oryu06}. For this reason, in subsequent calculations, the Coulomb matrix $T_C$ was always computed as the sum of only two terms, $T_{\phi}$ and $t^{R\phi}$. As direct calculations show, this sum is independent of the choice of screening parameter, provided condition (\ref{eq28}) is met, and for elastic scattering of charged particles, the result is exact.
	
	To account for the Coulomb interaction in short-range dynamics, the two-potential method is reapplied to the sum of the short-range potential $V_S$ (\ref{eq17},\ref{eq21}) and the Coulomb potential. As a result of incorporating the nuclear Coulomb interaction, the short-range $t$-matrix for single-channel scattering not only acquires a perturbation in the resolvent due to the appearance of the Coulomb $T_C$-matrix:
	\begin{equation}\label{eq33}
		t^{sR}=V_S+K^{sC}G_0t^{sR}\equiv V_S+V_S(1+G_0T_C)t^{sR},
	\end{equation}
	but also contains up to 37 terms in the full scattering $T$-matrix. However, on the mass shell, the number of terms reduces to four:
	\begin{equation}\label{eq34}
		T=(1+t_{R\phi}G_0)t^{sR}(1+G_0t_{R\phi}).
	\end{equation} 
	Expression (\ref{eq34}) is applicable to rearrangement and fusion reactions. For elastic scattering, an additional term in the form of the Coulomb $T_C$-matrix must be added to expression (\ref{eq34}).
	
	A specific mention should be made of the algorithm for calculating the initial state Coulomb interaction, which is relevant for fusion reactions such as $T(D,n)^4\text{He}$. In such reactions, the Coulomb interaction occurs only in the entrance channel, and thus must be accounted for as an ISI effect:
	\begin{equation}\label{eq35}
		T_{ISI}=T+TG_0T_C,
	\end{equation} 
	where $T$ is the short-range $t$-matrix calculated by formula (\ref{eq34}). Expression (\ref{eq35}) reflects the convergence to the mass shell for both the short-range $t$-matrix with respect to its right momentum and the Coulomb $T_C$-matrix with respect to its left momentum. Therefore, the use of formula (\ref{eq35}) is complicated by the necessity of pre-calculating short-range $T$-matrices, which already incorporate the Coulomb perturbation, across the entire range of kinetic energies considered. To address these technical challenges, a numerical computation technique was developed. This involves the preliminary calculation of Coulomb $T_C$-matrices, which are subsequently used in the next step to compute short-range $T$-matrices via formula (\ref{eq34}) for all relevant kinetic scattering energies. These data arrays are then substituted into the integral of formula (\ref{eq35}). Integration is performed using the stored data arrays for $T_C$ and $T$ as functions of their arguments, employing a criterion of maximum proximity between the stored arguments and the points of the numerical grid used for integration in (\ref{eq35}).
	
	Turning to the consideration of effects related to the off-shell behavior of the Coulomb $T_C$-matrix when accounting for the initial-state interaction (\ref{eq35}), it should be noted that, to simplify the calculations, the computation of $T_C$ was performed in an approximation where terms not equal to zero off the mass shell are omitted. For this reason, a qualitative estimation of the off-shell effect of the Coulomb interaction, which manifests only in one of the terms of the integral (\ref{eq35}) corresponding to a principal value integral, was performed by simply multiplying the obtained Coulomb $T_C$-matrix by a Gaussian damping factor:
	\begin{equation}\label{eq36}
		T_C^{\text{h.o.\,shell}}=T_C\times\exp{\big(-a(|p_i^2-p^2|)\big)}
	\end{equation}
	with a real controlling parameter $a>0$. In formula (\ref{eq36}), $p$ is the integration variable of (\ref{eq35}), and $p_i$ is the magnitude of the initial relative momentum of the colliding particles.

	\chapter{Results of total cross section calculations}

	\section{Coulomb contributions}
	
\begin{figure}
		\begin{center}
			\resizebox{0.8\textwidth}{!}{
				\includegraphics{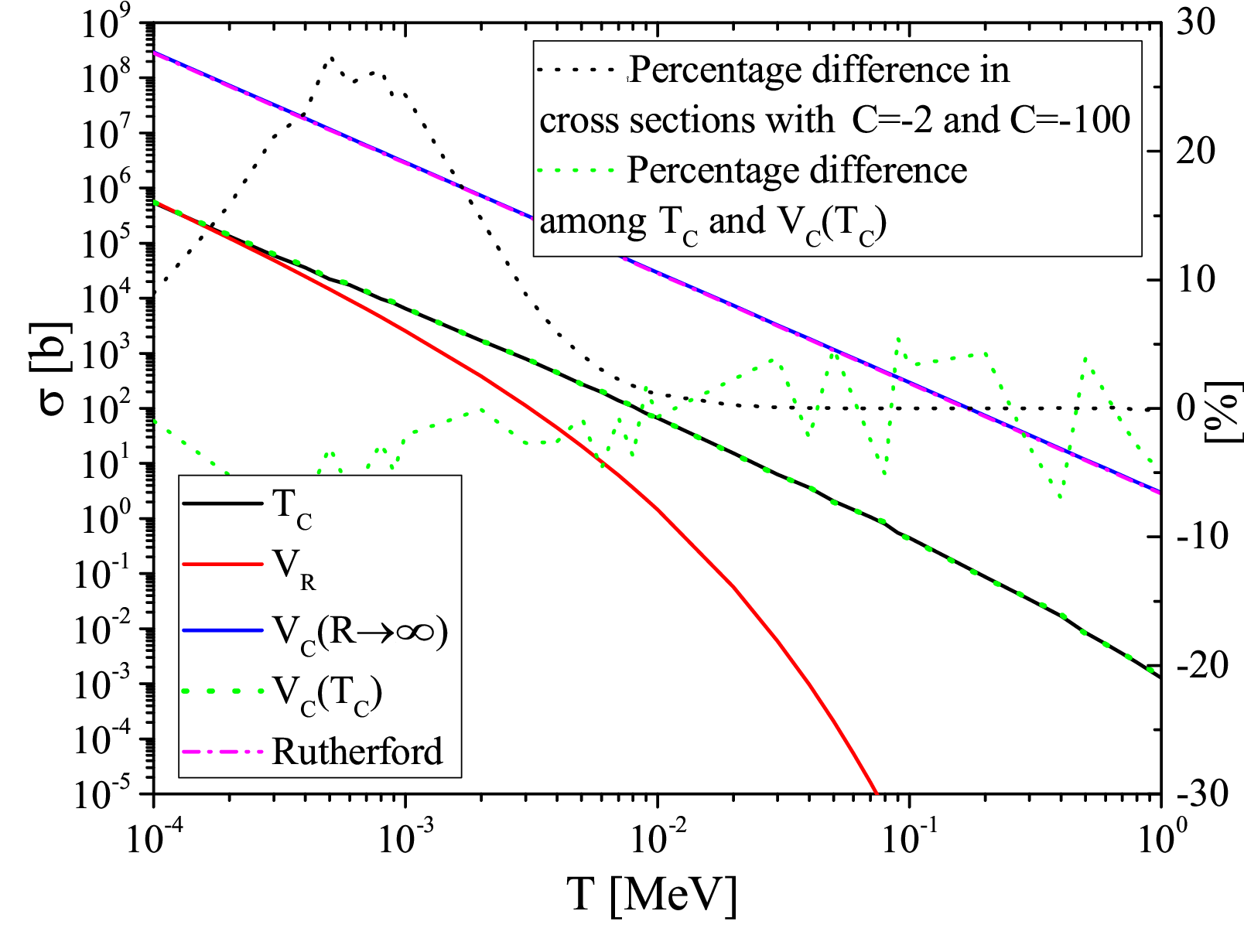}} 
			\caption{ Total cross section for Coulomb $DT$ scattering. Rutherford scattering (magenta dash-dotted line) is compared with contributions from the screened Coulomb potential $V_R$ and the unscreened Coulomb potential $V_C$ (in the limit $R\to\infty$). Calculations for the Coulomb $T_C$-matrix (\ref{eq29}) and the Coulomb potential $V_C(T_C)$ (obtained from the solution of equation (\ref{eq37})) are also presented. The right-hand scale shows the relative deviations (in percentages) between calculations performed with significantly different constants $C$ for the screening radius (\ref{eq27}).  }\label{fig1}
		\end{center}
	\end{figure}

	Coulomb electromagnetic contributions were calculated in three aspects, corresponding to points 1–3 of the previous section. The nuclear Coulomb interaction, exemplified by $DT$ scattering, was calculated using the two-potential formalism \cite{Oryu06}. The $T_C$-matrix obtained from formula (\ref{eq29}) was employed to calculate the elastic scattering cross section (\ref{eq7}). Similar calculations were performed for the screened potential $V_R$, using it instead of the $t$-matrix, as well as for the potential $V_C$ derived from the Lippmann-Schwinger equation:
	\begin{equation}\label{eq37}
		V_C(T_C):=(1+G_0T_C)^{-1}T_C
	\end{equation} 
	for a known $T_C$-matrix. The calculation results compared with the Rutherford formula are presented in Figure~(\ref{fig1}).
	%%%%%%%%%%%%%%%%%%FIGURE 1
	%%%%%%%%%%%%%%%%%END FIGURE 1
	It was of particular importance to demonstrate the independence of the $T_C$-matrix calculations using formula (\ref{eq29}) from the choice of parameter $C$ in the screening radius (\ref{eq27}). The right-hand scale of Figure~(\ref{fig1}) shows the relative percentage deviation between two cross section calculations with $T_C$-matrices for substantially different constants, $C = -2$ and $C = -100$. The discrepancy does not exceed 25$\%$ in the region $T \approx 1$~keV, and the differences become negligible as energy increases. For comparison, the right-hand scale of Figure~(\ref{fig1}) also shows the relative deviation between calculations using the exact Coulomb $T_C$-matrix and the potential determined by solving equation (\ref{eq37}). The differences are on the order of 5$\%$. It should be noted that these calculations were calibrated against the Rutherford scattering result, which is reproduced in the limit $R \to \infty$ using only the screened potential. The primary result of applying the two-potential formalism to account for nuclear Coulomb repulsion was a reduction in the cross section obtained using $T_C$ compared to Rutherford scattering. Calculations with the full Coulomb $T_C$-matrix approach those with the screened potential for a given screening radius (\ref{eq27}) only in the kinetic energy region $T \approx 0.1$~keV.
	
\begin{figure}
		\begin{center}
			\resizebox{1.0\textwidth}{!}{
				\includegraphics{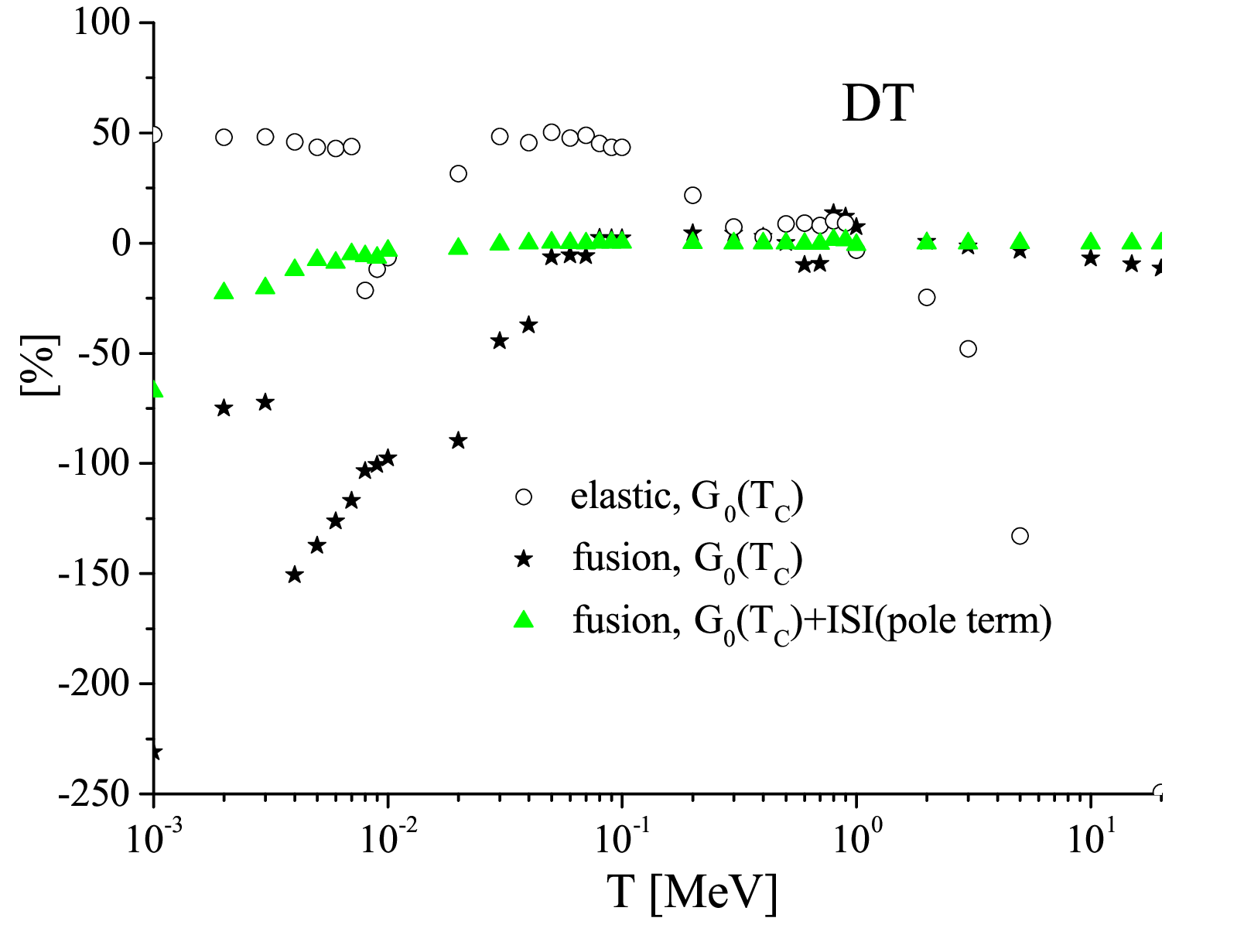}\hspace{-1cm}
				\includegraphics{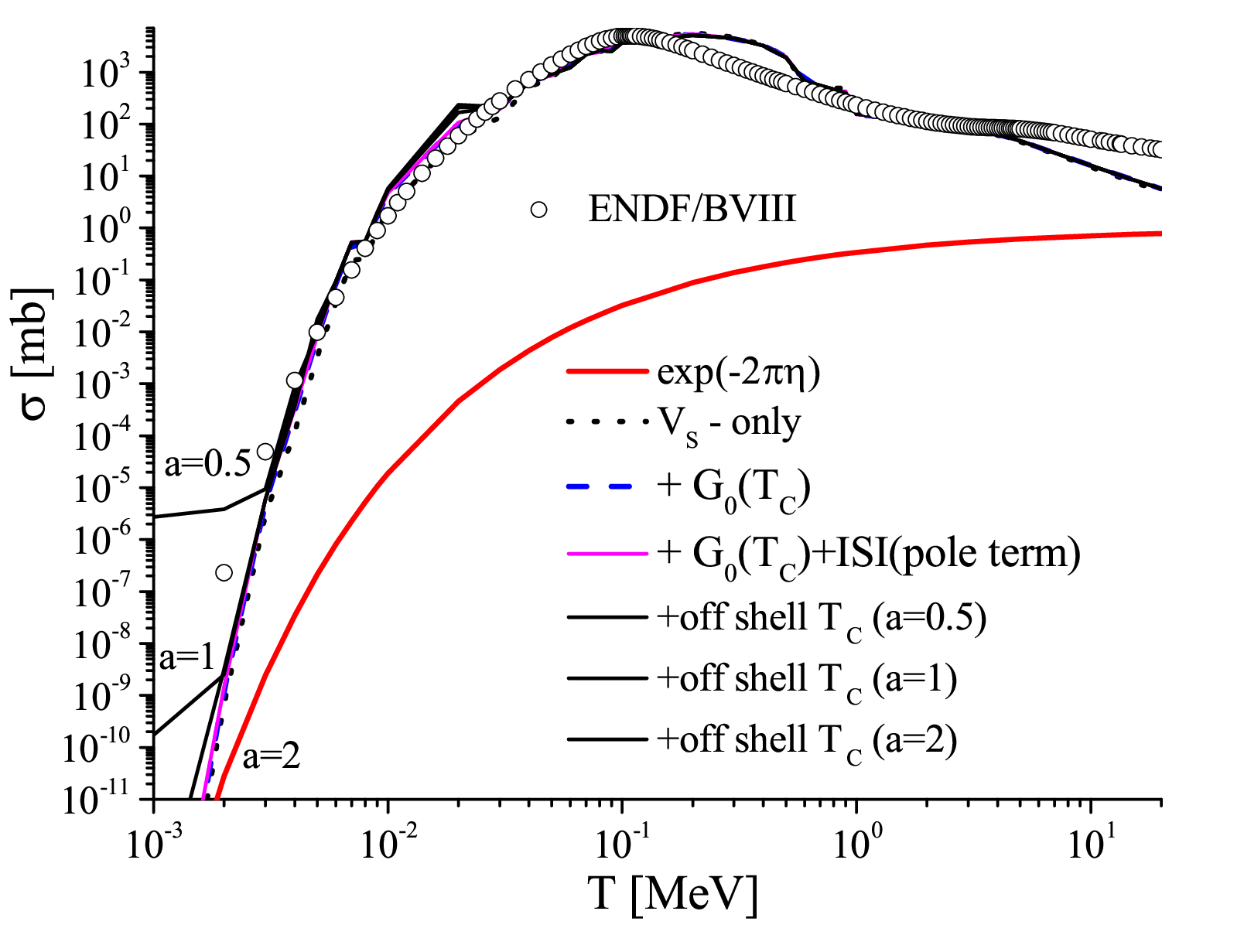}}
			\caption{ Left: Relative percentage contributions to the total cross section for elastic $DT$ scattering and the $T(D,n)^4$He reaction with a non-zero short-range interaction $V_S$ (\ref{eq17}), accounting for the Coulomb perturbation of the short-range resolvent $G_0(T_C)$, and with an additional inclusion of the pole contribution (\ref{eq35}) from the initial state interaction. Right: Total cross section for the $T(D,n)^4$He reaction obtained in this work with various Coulomb contributions, including estimates of off-shell Coulomb effects using formula (\ref{eq36}), compared with evaluated nuclear data from the ENDF library \cite{ENDF}. }\label{fig2}
		\end{center}
	\end{figure}

	In the subsequent stage, the nuclear Coulomb interaction model was integrated into the short-range dynamics. As previously noted, in this case, the Coulomb contributions manifest not only as a separate term for the $t$-matrix in elastic scattering but also in the short-range resolvent $G_0 \to G_0(T_C)$, as well as in the initial state interaction (ISI) for reactions of the $T(D,n)^4\text{He}$ type. Figure~(\ref{fig2})(left) shows the relative deviation in cross sections calculated with allowance for the Coulomb perturbation in the resolvents for elastic $DT$ scattering and the $T(D,n)^4\text{He}$ reaction. Additionally, for the reaction, initial state interaction calculations are presented, limited to the pole approximation in formula (\ref{eq35}). It can be seen that the Coulomb perturbations in the resolvent lead to an increase in the reaction cross section at $T < 100$~keV and a decrease in the elastic scattering cross section at $T < 300$~keV. Conversely, perturbations in the resolvent result in an increase in the elastic scattering cross section at $T > 2$~MeV. The pole contribution of the ISI effect, which is additional to the resolvent induced one, also leads to an increase in the reaction cross section in the region $T < 20$~keV. This effect is significant, reaching up to 67$\%$ at $T = 1$~keV. In elastic scattering, the Coulomb dynamics begin to dominate over the short-range dynamics already in the region $T < 350$~keV.
	%%%%%%%%%%%%%%%%%%FIGURE 2
	%%%%%%%%%%%%%%%%%END FIGURE 2
	An additional off-shell Coulomb effect, relative to those discussed above, associated with the convergence from the mass shell during integration in formula (\ref{eq35}), is presented in Figure~(\ref{fig2})(right). The figure shows calculations of the total cross section for the $T(D,n)^4\text{He}$ fusion reaction, obtained within the framework of the presented two-channel model based on the solution of coupled equations (\ref{eq20}). Good agreement is observed with the evaluated nuclear data from the ENDF library \cite{ENDF} across two orders of magnitude in energy, despite having only two controllable model parameters per channel, as reflected in Table~(\ref{tab1}). For illustration, the Gamow factor is also presented in Figure~(\ref{fig2})(right). A key factor with maximum influence on the total cross section of the $T(D,n)^4\text{He}$ reaction is the effect of the Coulomb $T_C$-matrix converging from the mass shell in the integral (\ref{eq35}). Cross section calculations were performed for three values of the controlling parameter $a$ in formula (\ref{eq36}): $a=0.5, 1, 2$. Since the lowest energy experimental point for the total cross section of the $T(D,n)^4\text{He}$ fusion reaction is located in the region $T \approx 10$~keV, the observed contribution from the Coulomb $T_C$-matrix converging from the mass shell may be of experimental interest, despite the relatively small cross sections (on the order of nanobarns).
	
\begin{figure}
		\begin{center}
			\resizebox{0.8\textwidth}{!}{
				\includegraphics{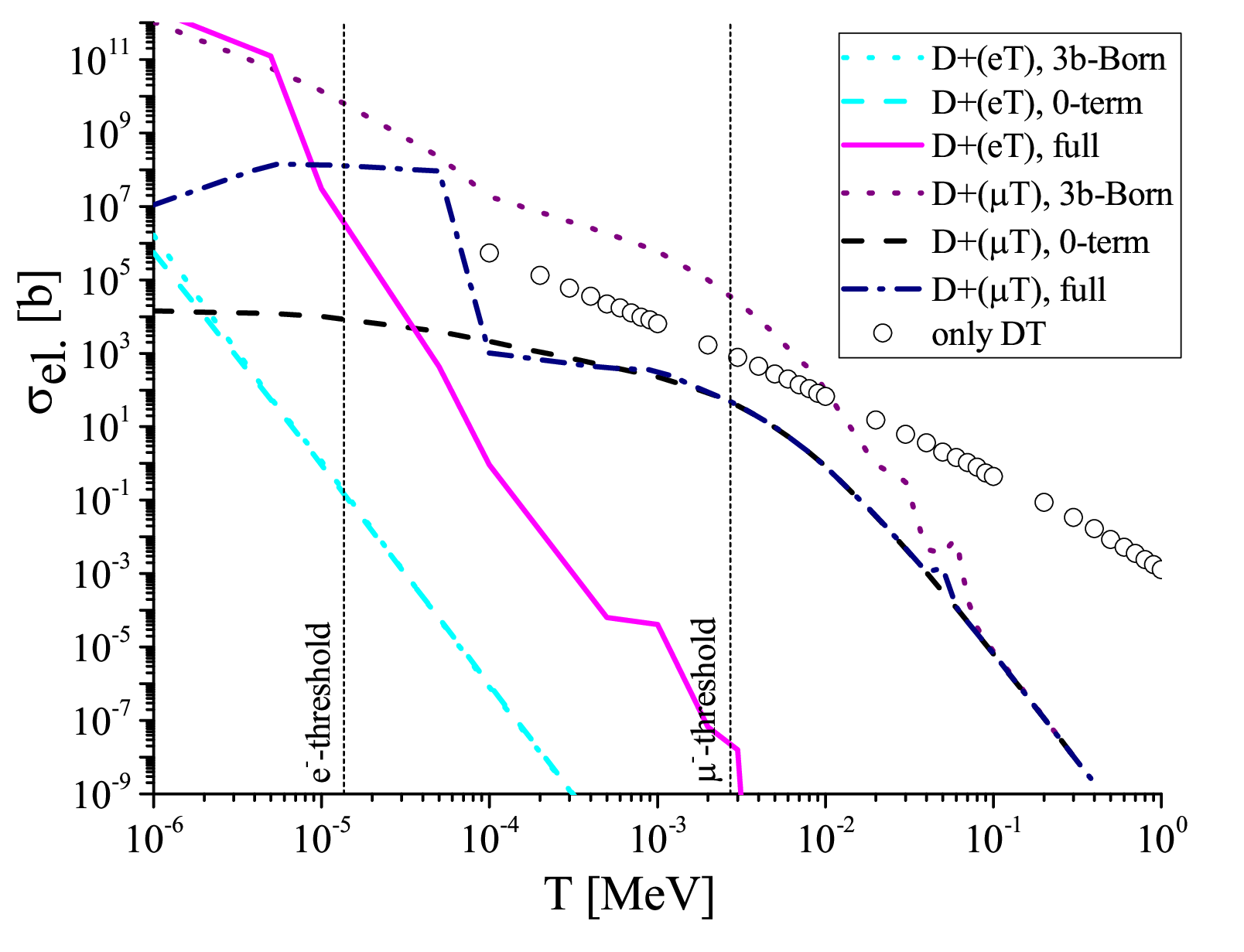}} 
			\caption{Total cross section for the elastic Coulomb scattering of deuterons by $eT$ and $\mu T$ atoms in the three-body Born approximation (3b-Born using formula (\ref{eq38})) and according to formula (\ref{eq11}) (full), including the case where integral terms are neglected (0-term). For comparison, pure Coulomb $DT$ scattering calculations are presented, and the ionization thresholds for the electron ($eT$) and muonic ($\mu T$) atoms are indicated by short vertical dashes for clarity. }\label{fig3}
		\end{center}
	\end{figure}

	In the next stage of assessing Coulomb contributions, anti-screening effects induced by atomic electrons were included in the consideration. Specifically, for elastic $DT$ scattering, the influence of the target atomic electron ($eT$) was evaluated within a three-body model based on the Faddeev equation for the three-body $T$-matrix, including only one ground atomic state of the target. Calculations were performed using the formula for the elastic scattering amplitude (\ref{eq11}), and also neglecting the integral contributions in this formula (the result denoted as <<0-term>>). The calculated cross section for elastic $D-(eT)$ scattering is presented in Figure~(\ref{fig3}). 
	%%%%%%%%%%%%%%%%%%FIGURE 3
	%%%%%%%%%%%%%%%%%END FIGURE 3
	It should be noted that the three-body Faddeev equations in this two-potential approach were not modified as prescribed by rigorous theory \cite{Oryu06}. The Faddeev equations themselves were solved iteratively, identifying regions of logarithmic singularities for each iteration and interpolating these regions using cubic splines for the divergent logarithm (\ref{eq14}). Thus, the $T_C$ matrices for nucleus-nucleus and lepton-nucleus interactions found in the previous stage in the two-body sectors were incorporated into the Faddeev equations without additional modifications that would increase the number of integrals and integral kernels of the solved Faddeev equations.
	For comparison, calculations using the three-body Born approximation are presented in Figure~(\ref{fig3}) according to the formula:
	\begin{align}\label{eq38}
		\begin{gathered}
			\langle \Psi_{1(23)}| U_{11}| \Psi_{1(23)}\rangle =
			\phi_{(23)}\big(-{\bf q}\tfrac{m_1}{m_1+m_3}-{\bf q}_0\big)\big(E-\tfrac{q^2}{2\mu_{23}}-\tfrac{q_0^2}{2\mu_{13}}-
			\tfrac{q_0q y_{qq_0}}{m_3} \big)\times \\
			\phi_{(31)}\big({\bf q}+{\bf q}_0\tfrac{m_2}{m_2+m_3}\big)  \\
			+
			\phi_{(23)}\big({\bf q}\tfrac{m_1}{m_1+m_2}+{\bf q}_0\big)\big(E-\tfrac{q^2}{2\mu_{23}}-\tfrac{q_0^2}{2\mu_{12}}-
			\tfrac{q_0q y_{qq_0}}{m_2} \big)\phi_{(12)}\big(-{\bf q}-{\bf q}_0\tfrac{m_3}{m_2+m_3}\big)+\\
			\int d^3 q^{'}\Big[
			\phi_{(23)}\big(-{\bf q}\tfrac{m_1}{m_1+m_3}-{\bf q}'\big) V_R\big({\bf q}+{\bf q}'\tfrac{m_2}{m_2+m_3},-{\bf q}_0-{\bf q}\tfrac{m_1}{m_1+m_3}\big)\times  \\
			\phi_{(23)}\big({\bf q}+{\bf q}_0\tfrac{m_2}{m_2+m_3}\big) 
			+   
			\phi_{(23)}\big({\bf q}\tfrac{m_1}{m_1+m_3}+{\bf q}'\big) V_R\big(-{\bf q}-{\bf q}'\tfrac{m_3}{m_2+m_3},{\bf q}_0+{\bf q}\tfrac{m_1}{m_1+m_2}\big)\times  \\
			\phi_{(23)}\big(-{\bf q}-{\bf q}_0\tfrac{m_3}{m_2+m_3}\big) 
			\Big],  
		\end{gathered}
	\end{align}
	which is obtained from the amplitude (\ref{eq11}) through a recursive series of simplifications $T\to t\to V_R$, wherein the three-body $T$-matrices in the integral kernels are replaced by screened Coulomb potentials $V_R$. The result of the calculation according to formula (\ref{eq38}) in this approximation is close to the 0-term calculations (neglecting the integrals) in the region $T<10$~eV for elastic $D+(eT)$ scattering. It was of interest to illustrate the behavior of three-body calculations of electron anti-screening of the nuclear Coulomb interaction by simply replacing the electron mass with the muon mass, while keeping all other parameters constant. The result of such a calculation demonstrates a characteristic convergence of the full calculation (full) based on formula (\ref{eq11}) toward the 0-term approximation already in the $T>100$~eV region. Both calculations exhibit a characteristic kink in the cross section in the vicinity of the $\mu^-$--ionization threshold. The latter is attributed to the growth of the cross section itself, information about which is also encapsulated in the solved system of Faddeev equations. However, in the case of scattering on a muonic atom, the Born approximation (\ref{eq38}) exceeds the full calculation via formula (\ref{eq11}) over a wide energy range $T<100$~keV. For elastic $D+(eT)$ scattering, the total cross section is higher than that of Coulomb $DT$ scattering in the region below the ionization threshold, but subsequently decreases more rapidly with increasing energy than the Coulomb $DT$ interaction cross section.
	
	Since the elastic $D+(eT)$ scattering cross section exhibits a sharp drop in the region of the $e^-$-ionization threshold, the electron leaving the interaction zone can no longer exert an anti-screening influence on the progression of nuclear reactions between the nuclei. Consequently, in further cross section calculations using a cluster representation of the target nucleus, the electron anti-screening effect will not be taken into account. The fact that the ionization threshold for a muonic atom lies at relatively high energies implies that the anti-screening role of the muon may manifest at energies as low as $T\approx 3$~keV, leading to a sharp increase in the fusion cross section.

	\section{Total reaction cross sections in the cluster representation}

	The fusion reactions $^3\text{He}(T,np)^4\text{He}$, $^3\text{He}(T,D)^4\text{He}$, and $^3\text{He}(^3\text{He},2p)^4\text{He}$ serve as transitional processes between conventional fusion reactions involving hydrogen isotopes and the fusion of light nuclei with charge greater than unity and producing more than two particles. In this context, these reactions offer a unique opportunity to quantify the magnitude of Coulomb repulsion across a broad energy range and to corroborate the cluster reaction mechanism within the framework of three-body dynamics.
	
\begin{figure}
		\begin{center}
			\resizebox{1.0\textwidth}{!}{
				\includegraphics{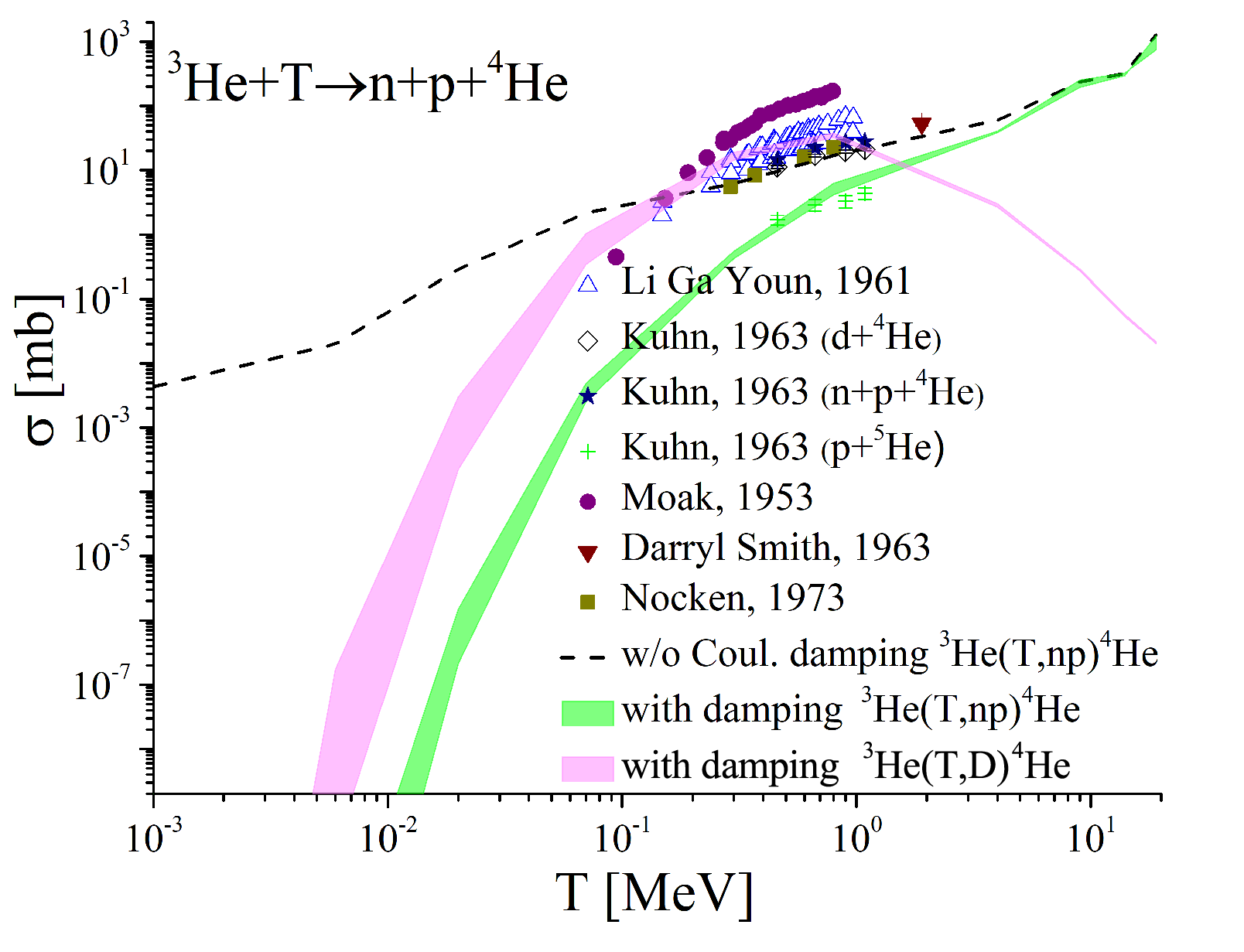}\hspace{-1cm}
				\includegraphics{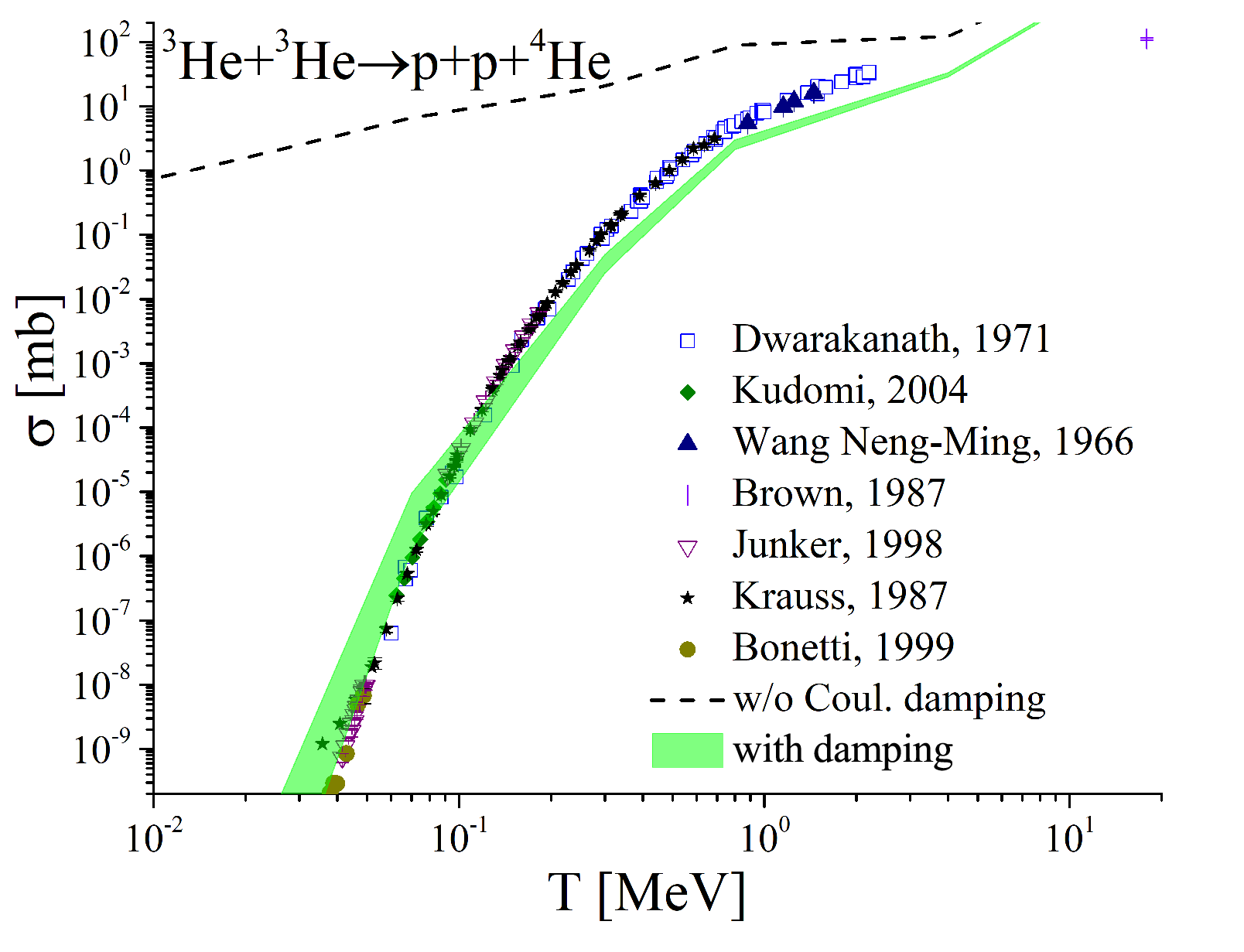}}
			\caption{Left panel: Total cross sections for the reactions $^3\text{He}(T,np)^4\text{He}$ and $^3\text{He}(T,D)^4\text{He}$ obtained without considering Coulomb repulsion (dashed line for the $^3\text{He}(T,np)^4\text{He}$ process) and with Coulomb repulsion taken into account, where results were obtained both by employing the Gamow factor (lower bound of the shaded area) and by using the square of function (\ref{eq23}) with a non-trivial angular dependence and a screening radius $r\equiv R$ from formula (\ref{eq27}). Experimental data are from \cite{Youn,Kuhn,Moak,Smith,Nocken} Right panel: Analogous calculations of the total cross section for the $^3\text{He}(^3\text{He},2p)^4\text{He}$ reaction. Experimental data are from \cite{Dwarakanath,Kudomi,Ming,Brown,Junker,Krauss,Bonetti}.  }\label{fig4}
		\end{center}
	\end{figure}

	As previously noted, the fusion of $^3\text{He}$ and $T$ nuclei is understood within a cluster representation, where two distinct fusion mechanisms, namely $DT$ and $D^3\text{He}$ , provide coherent contributions to the resultant scattering matrix. Crucially, the third spectator particle (a proton and a neutron, respectively) does not participate directly in the fusion process. Consequently, the coupling between rearrangement channels and the three-body channels  $np^4\text{He}$ and $nD^3\text{He}$  is established within the identical system of Faddeev integral equations (\ref{eq12}), thereby obviating the need for supplementary model considerations or simplifications. This methodology, employing Faddeev equations for the simultaneous description of elastic scattering, fusion reactions, and reactions involving the ejection of all three particles into the energy continuum, represents a novel approach. A comparative analysis of the cross sections predicted by cluster theory with extant experimental data permits inferences regarding the contributions of individual cluster configurations to the mechanism of direct fusion reactions. A comparison with the experimental data presented in Figure~(\ref{fig4}) for the total cross sections of the reactions $^3\text{He}(T,np)^4\text{He}$, $^3\text{He}(T,D)^4\text{He}$, and $^3\text{He}(^3\text{He},2p)^4\text{He}$ corroborates that the developed multicluster approach for describing direct reaction mechanisms is indeed pertinent and accounts for a substantial portion of the overall reaction cross section.
	%%%%%%%%%%%%%%%%%%FIGURE 4
	%%%%%%%%%%%%%%%%%END FIGURE 4
	The assessment of Coulomb repulsion effects between colliding nuclei in the $^3\text{He}(T,np)^4\text{He}$, $^3\text{He}(T,D)^4\text{He}$, and $^3\text{He}(^3\text{He},2p)^4\text{He}$ reactions was conducted utilizing both the conventional Gamow factor and an explicit form of the scattering state  (\ref{eq23}), where the parameter $r=R$ corresponds to the screening radius defined by (\ref{eq27}). This particular choice of parameter precludes the divergence of the hypergeometric function with increasing relative momentum. Concurrently, it establishes a connection between the screening zone $R$ in the two-potential method and the purely Coulomb scattering state (\ref{eq23}), which inherently exhibits coordinate dependence. In contrast to the previously analyzed $T(D,n)^4$He fusion reaction, for helium-3 nucleus fusion, the explicit inclusion of the angular dependence of function (\ref{eq23}) becomes manifest at energies as low as $T<100$~keV. While this effect is inconsequential for describing nuclear reaction cross sections, it proves crucial for the accurate characterization of astrophysical $S$-factors and for the conversion of total cross sections $S(E)\to\sigma(E)$  when relying solely on the Gamow factor.
	
	It is noteworthy to observe the consistency between the calculated cross sections for the  $^3\text{He}(T,np)^4\text{He}$ and $^3\text{He}(T,D)^4\text{He}$ processes and the somewhat contradictory experimental data. The theoretical framework presented herein predicts a more substantial contribution from the $^3\text{He}(T,D)^4\text{He}$ rearrangement channel. This contribution, while interpolating between the experimental data points from \cite{Youn,Kuhn}, exhibits a sharp decline in the energy regime above $T>1$~MeV. Concurrently, the calculated cross sections for $^3\text{He}(T,np)^4\text{He}$ demonstrate good agreement with the data from \cite{Kuhn}, specifically those reported for the $p^5\text{He}$ channel. These cross sections subsequently surpass those of the rearrangement channel in the $T>1$~MeV energy range. It is crucial to emphasize that the predominant influence of Coulomb repulsion between colliding nuclei plays a discernible role across the entire investigated energy range of $T\in[1~\text{keV},20~\text{MeV}]$. Numerical data pertaining to all discussed fusion and rearrangement reactions are provided in Table~(\ref{tab2}) for selected energies within this range.
	%%%%%%%%%%%%%%%%%%%%%%%%%%%%%%%%%%%%%%%TABLE2%%%%%%%%%%
	\begin{table}[p]
		\caption{Calculated total cross sections [in millibarns] for the considered processes in the cluster representation of nuclei within a three-body dynamics framework.}\label{tab2}	
		\bigskip
		\begin{tabular}{|c|c|c|c|c|c|c|c|}
		%\begin{tabular*}{\tblwidth}{@{}LLLLLLLL@{}}
			\hline
			$T$~[MeV]  & 0.07  & 0.3  & 0.8 & 4  & 9 & 14 & 19 \\
			\hline
			$^3\text{He}(T,T)^3\text{He}$ & 0.52  & 17.3 & 16.3 & 0.3 & 0.06 & 0.02 & 0.015 \\
			$^3\text{He}(T,D)^4\text{He}$ & 0.34 & 14.4 & 32.3 & 2.7 & 0.27 & 0.054 & 0.02 \\
			$^3\text{He}(T,np)^4\text{He}$  & 0.003 & 0.42 & 4.15 & 38.5 & 195 & 294 & 1257 \\
			$^3\text{He}(T,nD)^3\text{He}$ &  -  &   - & - & - & - & 9.37 & 28.8 \\
			\hline 
			$^3\text{He}(^3\text{He},2p)^4\text{He}$ & 1.3$\cdot 10^{-6}$ & 0.03 & 2.13 & 28.9 & 283 & 371 & 1055 \\
			$^3\text{He}(^3\text{He},pD)^3\text{He}$  & - & - & - & - & - & 11.5 & 25.2 \\
			\hline
			$^7\text{Li}(^3\text{He},\phantom{0}^3\text{He})^7\text{Li}$ &
			4.4$\cdot 10^{-4}$ & 1.3$\cdot 10^{-2}$ &  2.5$\cdot 10^{-2}$ & 9.3  & 0.45 & 11.8 & 0.63  \\
			$^7\text{Li}(^3\text{He},\phantom{0}^4\text{He})^6\text{Li}$  &
			1.33$\cdot 10^{-2}$  & 0.35 & 1.61 & 96.6 & 13.8 & 50.3 & 13.9 \\
			$^7\text{Li}(^3\text{He},D^4\text{He})^4\text{He}$  & 1.5$\cdot 10^{-3}$  & 0.16 & 0.58 & 41.2 & 698 & 2722 & 3210  \\
			$^7\text{Li}(^3\text{He},T^3\text{He})^4\text{He}$ & - & - & - & 0.83 & 9.88 & 31.7 & 82.6 \\ 
			\hline
		\end{tabular}
	\end{table}
	%%%%%%%%%%%%%%%%%%%%%%%END TABLE2%%%%%%%%%%%%%%%%%%
	Given the good agreement of the calculated cross sections for fusion and rearrangement reactions involving helium isotopes, it is of interest to consider the following fusion reactions of helium-3 isotopes with lithium-7 type nuclei, which also admit a cluster description in their ground state. Within this cluster description, the $^7$Li nucleus can be represented as a bound $T-^4$He pair in a $J^P=3/2^-$ $p$-wave state. In contrast to the fusion of helium and hydrogen isotopes, there are scarcely any experimental data available for the total cross sections of $^3$He nuclei fusion with lithium isotopes. Only the experimental work \cite{Forsyth} on the fusion cross sections for the $^7\text{Li}(^3\text{He},\phantom{0}^4\text{He})^6\text{Li}$ reaction and optical model calculations of the elastic scattering cross section for $^3\text{He}^6\text{Li}$ nuclei \cite{Ludecke} are available.
	
	For the cluster description of the fusion reactions $^7\text{Li}(^3\text{He},\phantom{0}^4\text{He})^6\text{Li}$, $^7\text{Li}(^3\text{He},D^4\text{He})^4\text{He}$ and the breakup reaction $^7\text{Li}(^3\text{He},T^3\text{He})^4\text{He}$, separable potentials  (\ref{eq21}) were employed for the elastic $T^4\text{He}$ and $^4\text{He}^4\text{He}$ interactions, with parameters presented in Table~(\ref{tab1}). Additionally, parameters for elastic $D^4\text{He}$ scattering from \cite{EgorovnLi} were utilized, with an interaction strength $\lambda$ calculated from the condition:
	\begin{equation}\label{eq39}
		\lambda^{-1}=\frac{\mu}{\pi^2}\int\limits_0^{\infty}\frac{\xi^2(p)p^2 dp}{2\mu|E_b|+p^2},
	\end{equation}
	with a binding energy $E_b$ of the $D-^4$He clusters in the ground state of the $^6$Li nucleus.
	
\begin{figure}
		\begin{center}
			\resizebox{0.8\textwidth}{!}{
				\includegraphics{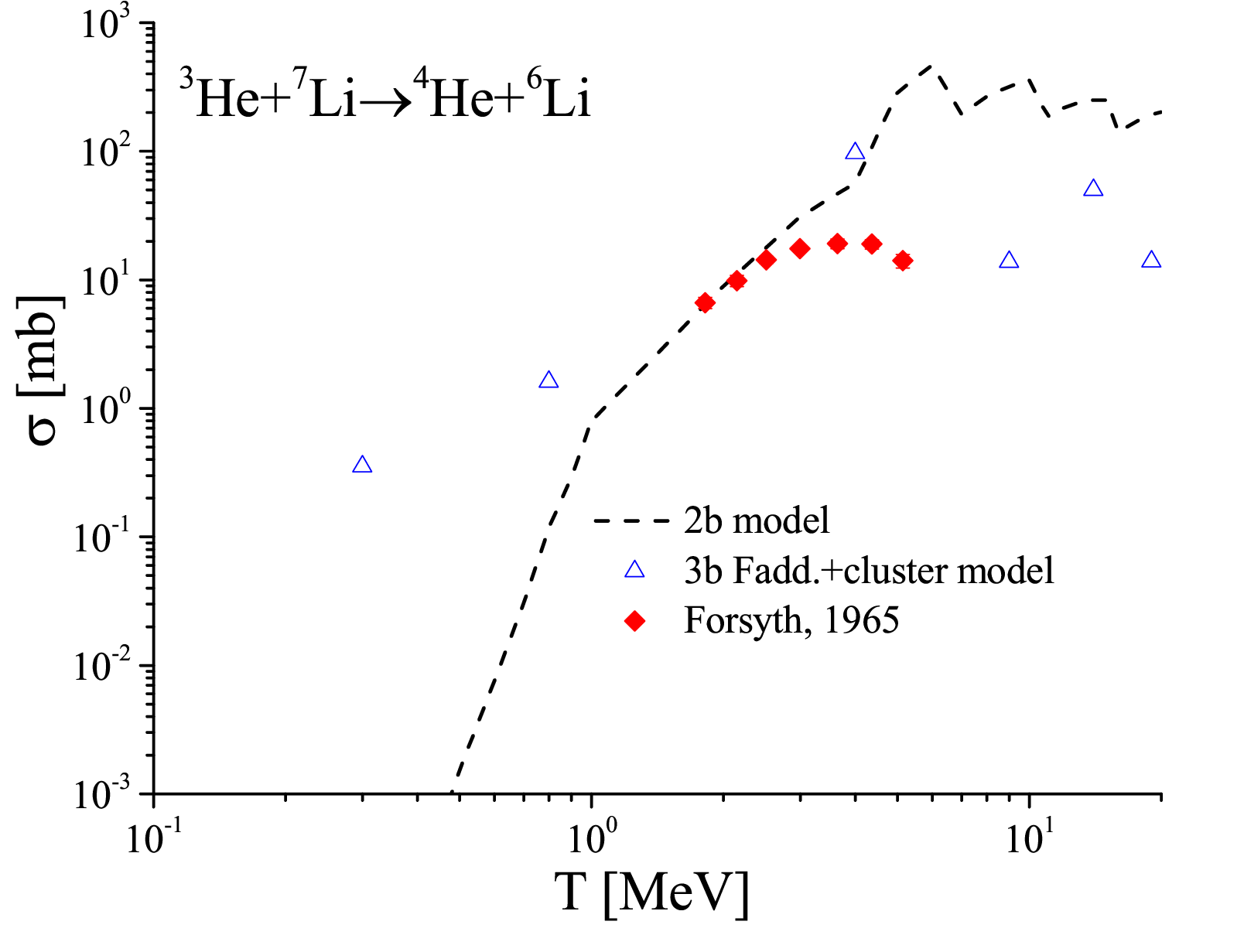}} 
			\caption{Total cross section for the  $^7\text{Li}(^3\text{He},\phantom{0}^4\text{He})^6\text{Li}$  reaction, calculated using two-channel coupled Lippmann-Schwinger equations (\ref{eq20}) and based on three-body Faddeev dynamics with the target nucleus represented as a $^4\text{He}-T$ cluster system in a $J^P=3/2^-$ state in the $L=1$ partial wave. Experimental data for the ground state of $^6$Li are from \cite{Forsyth}. }\label{fig5}
		\end{center}
	\end{figure}

	Furthermore, the cluster description of these $^3$He and $^7$Li fusion reactions necessitates the input of $t$--matrices for the $^3\text{He}(T,D)^4$He reaction and the elastic scattering $^3\text{He}(T,T)^3$He, which were determined in a preceding stage within the framework of the corresponding system of Faddeev equations (\ref{eq12}). Such an application of the Faddeev equations, based on the principle of embedding one scattering problem as a subproblem within another, enhances the overall predictive power of the developed multicluster approach and minimizes the number of unknown parameters (ingredients) of the model. The calculated total cross section for the $^7\text{Li}(^3\text{He},\phantom{0}^4\text{He})^6\text{Li}$ reaction is presented in Figure~(\ref{fig5}).
	%%%%%%%%%%%%%%%%%%FIGURE 5
	%%%%%%%%%%%%%%%%%END FIGURE 5
	In order to compare with three-body calculations, which did not require any {\it a priori} information regarding reactions between  $^3$He and  $^7$Li nuclei, a simple two-body potential model featuring a local potential (\ref{eq17}) was developed. This model incorporates the coupling of only two elastic scattering channels and the $^7\text{Li}(^3\text{He},\phantom{0}^4\text{He})^6\text{Li}$ reaction channel. The parameters of this simplified model were adjusted solely to reproduce the total cross sections of elastic $^3\text{He}^6\text{Li}$ scattering, as obtained from optical model calculations by \cite{Ludecke}. Consequently, the reaction channel in this model was not calibrated. Nevertheless, these simplified calculations show satisfactory agreement with experimental data \cite{Forsyth} in the energy range $T<2.5$~MeV, though they substantially overestimate it for $T>4$~MeV. Conversely, calculations of the same reaction cross section within the three-body multicluster framework demonstrate agreement with experiment in the range $T\approx 1-2$~MeV, exhibiting a smooth decrease of the cross section with diminishing energy.
	
	It is noteworthy that the three-body calculation of the $^7\text{Li}(^3\text{He},\phantom{0}^4\text{He})^6\text{Li}$  cross section exhibits a localized jump at $T=4$~MeV, which may be attributed to numerical artifacts. Excluding this particular point, the agreement between the multicluster theory and experiment becomes acceptable. A comprehensive list of scattering processes and reactions between  $^3$He and $^7$Li nuclei investigated within the three-body Faddeev dynamics in the multicluster approach, for specific numerical energy values, is provided in Table~(\ref{tab2}). At the aforementioned level, the utilization of the cluster structure of light nuclei and the cluster reaction mechanism where fusion and rearrangement processes primarily occur between individual clusters rather than nucleons within a few-body dynamical framework relying on Faddeev integral equations, leads to the description of the vast majority of experimental total cross sections for all examined reactions.
	
	Regarding elastic Coulomb scattering for  $^3$He$^7$Li  and $^3$He+(e$^7$Li) within a simplified local potential model, an assessment of electron anti-screening reveals a more rapid decay of the $^3$He+(e$^7$Li) cross section compared to $^3$He$^7$Li above the ionization threshold. This observation further indicates the negligible influence of the electron on the elastic scattering process and, consequently, on the fusion reaction rate. The incorporation of the three-body Coulomb scattering amplitude, derived from equation (\ref{eq11}), into the integral (\ref{eq35}) for calculating the Coulomb ISI effect in the $^7\text{Li}(^3\text{He},\phantom{0}^4\text{He})^6\text{Li}$  reaction, within a basic two-channel potential model, yields several consequences. Firstly, for energies $T>6$~keV, the parameter $a$, which governs the convergence of the $T_C$-matrix to the mass shell, ceases to play a significant role. Secondly, the three-body Coulomb effect for $^3$He+(e$^7$Li) surpasses the Coulomb repulsion for  $^3$He$^7$Li by approximately 20$\%$.   Finally, at low energies, the dominant Coulomb contribution arises from the initial state interaction, whereas at high energies, it originates from the Coulomb perturbation within the short-range resolvent.
	
	The reported calculated fusion reaction cross sections, coupled with the conducted assessments of Coulomb effects and an accurate treatment of three-body dynamics within a multicluster framework for reactions across all channels accessible at energies $T<20$~MeV, constitute a significant advancement towards understanding the mechanisms of direct and rearrangement reactions occurring between light nuclei. The implemented approach, employing a cluster representation of both the target and incident nuclei within exact three-body dynamical methods, has demonstrated, through comparison with experimental data, high predictive power and confirmed the applicability of the utilized cluster expansions.

	\chapter{Conclusion}

	This work implements the application of few-body dynamics methods, based on Faddeev integral equations in momentum space, to determine the total cross sections for fusion and breakup reactions involving the emission of two and three particles into the continuum, utilizing a cluster representation of the colliding nuclei. A system of Faddeev integral equations for the three-body $T$-matrices of a system, which permits a nuclear reaction with particle type rearrangement in one of its channels, was derived and correctly solved. The total cross sections were obtained for the reactions $^3\text{He}(T,D)^4\text{He}$, $^3\text{He}(T,np)^4\text{He}$, $^3\text{He}(T,nD)^3\text{He}$,
	$^3\text{He}(^3\text{He},2p)^4\text{He}$,   $^3\text{He}(^3\text{He},pD)^3\text{He}$,
	$^7\text{Li}(^3\text{He},\phantom{0}^4\text{He})^6\text{Li}$, 	  $^7\text{Li}(^3\text{He},D^4\text{He})^4\text{He}$, and
	$^7\text{Li}(^3\text{He},T^3\text{He})^4\text{He}$, in which both the incident particle and the target nucleus were considered in a cluster representation. A two-potential method was implemented in this work for determining the Coulomb $t$-matrix and for accounting for Coulomb effects in the short-range dynamics within momentum space. Calculations of the initial state Coulomb interaction were performed, an estimation of the off-shell convergence effect for the Coulomb $t$-matrix was obtained, and the magnitude of the anti-screening effect due to atomic electrons on the Coulomb interaction of colliding nuclei was determined. Despite the fact that the magnitude of these effects remains small and not discernible in the total cross section of the studied fusion reactions, it may be of interest for astrophysical applications.
	
	From the comparison of the obtained calculated total cross sections for fusion reactions with available experimental data, two main results can be highlighted: the cluster representation for the mechanism of the investigated fusion reactions accounts for the major part of the total cross section in the range $T\in[1~\text{keV},20~\text{MeV}]$, and throughout this entire range, the Coulomb repulsion between colliding nuclei plays a significant role. Furthermore, the maximum suppression effect on the cross section due to Coulomb repulsion is realized through the Gamow factor. Subsequently, the presented multicluster approach for determining total cross sections can be applied to describe differential cross sections, as well as to investigate the cluster mechanism in other fusion reactions by employing more refined potentials or pairwise scattering matrices.

	\section*{Acknowledgments}
	
	The author acknowledges the financial support from the BASIS Foundation for the Development of Theoretical Physics and Mathematics (Project No. 23-1-3-3-1).
	
	The research was carried out with the support of a grant from the Government of the Russian Federation (Agreement No. 075-15-2025-009 of 28 February 2025)


\begin{thebibliography}{99}
		\bibitem{IAEA2024}
		A. Anikeev, V. Artisiuk, M.  Ascic, S. Ashraf, M.  Barbarino, J. Barton, C. Bellehumeur,  G. Catena, W. Chae Kim, C. Cheong, J. Donovan, G. Federici, M. Finnerty, N. Ganzarski, A. Goodman, A. Jasper, {\it et al}, IAEA World Fusion Outlook 2024 [electronic source]:
		https://www.iaea.org/publications/15777/iaea-world-fusion-outlook-2024
		\bibitem{Peni2021}
		Yu. E. Penionzhkevich, R. G.  Kalpakchieva, R. G.,
		{\it Light Exotic Nuclei Near the Boundary of Neutron Stability},
		(World Scientific Publ., 2021) 
		\bibitem{CDCC22}
		K. Hagino, K.  Ogata, A. M. Moro,
		Coupled-channels calculations for nuclear reactions: From
		exotic nuclei to superheavy elements,
		Prog. Part. Nucl. Phys. {\bf 65}, 103951 (2022) 
		\bibitem{SANCSM}
		K. D. Launey, G. H.  Sargsyan, A.  Mercenne, J. E. Escher, D. C.  Mumma,
		Ab initio symmetry-adapted approaches to nuclear reactions,
		Prog. Part. Nucl. Phys. {\bf148}, 104233  (2026)
		\bibitem{GloeckleCoul}
		W. Gl\"ockle,  J.  Golak,  R.  Skibi\'nski,  H.  Wita\l a H.,
		Exact three-dimensional wave function and the on-shell t-matrix for the
		sharply cut off Coulomb potential: failure of the standard renormalization
		factor, Phys. Rev. C {\bf 79}, 044003 (2009) 
		\bibitem{Kouzakov}
		K. Kouzakov, Yu. V. Popov,  V. L. Shablov,
		Comment on <<Exact three-dimensional wave function and the on-shell t-matrix for the sharply cut-off 
		Coulomb potential: Failure of the standard renormalization factor>>,
		Phys. Rev. C {\bf 81}, 019801 (2010)
		\bibitem{Deltuva}
		A. Deltuva, A. C. Fonseca,  P. U.  Sauer,
		Comment on <<Exact three-dimensional wave function and the on-shell t-matrix for the sharply cut-off
		Coulomb potential: Failure of the standard renormalization factor>>,
		Phys. Rev. C {\bf 81}, 019802 (2010)
		\bibitem{Aliotta}
		M. Aliotta, F.  Raiola,  G.  Gy\"urky,  A.  Formicola, R.  Bonetti,  C. Broggini,  L. Campajola,  P. Corisiero, H. Costantini, A. D'Onofrio, Z. F\"ul\"op, G. Gervino, L. Gialanella, A. Gudlielmetti, C. Gustavino, G. Imbriani, {\it et al.},
		Electron screening effect in the reactions $^3$He(d, p)$^4$He and d($^3$He, p)$^4$He,
		Nucl. Phys. A  {\bf 690}, 790 (2001) 
		\bibitem{Engstler}
		S. Engstler, A.  Krauss, K.  Neldner,  C. Rolfs,  U.  Schr\"oder,  K.  Langanke,
		Effects of electron screening on the $^3$He(d, p)$^4$He low-energy cross sections,
		Phys.  Lett. B {\bf 202}, 179 (1988) 
		\bibitem{Pratti}
		P. Pratti, C. Arpesella,  F.   Bartolucci, H. W. Becker,  E.  Bellotti,  C. Broggini, P. Corvisiero, G. Fiorentino, A. Fubini, G. Gervino, F. Gorris, U. Greife, C. Gustavino, M. Junker, C. Rolfs, W. H. Schulte, {\it et al.},
		Electron screening in the d+$^3$He fusion reaction, Zeitschrift f\"ur Physik A {\bf 350}, 171 (1994)
		\bibitem{Bracci}
		L. Bracci, G.  Fiorentini,  V. S.  Melezhik,  G.  Mezzorani, P. Quarati,
		Atomic effects in the determination of nuclear cross sections of astrophysical interest,
		Nucl. Phys. A {\bf 513}, 316 (1990)
		\bibitem{Fadd}
		L. D., Faddeev,  Scattering theory for a three-particle system, Sov. Phys. JETP, {\bf 12}, 1014 (1961)
		\bibitem{Yakub}
		O. A. Yakubovsky, On the Integral equations in the theory of N particle scattering, Sov.  J. Nucl. Phys. {\bf 5}, 937 (1967)   
		\bibitem{Merk}
		S. P. Merkuriev, On the Three-body Coulomb scattering problem, Annals of Physics {\bf 130}, 395 (1980).
		\bibitem{Lombardo}
		I. Lombardo, D. Dell'Aquila, Clusters in light nuclei: history and recent developments, La Rivista del Nuovo Cimento {\bf 46}, 521 (2023).
		\bibitem{Perrotta}
		S. S. Perrotta, M. Colonna, J. A. Lay, Clustering effects in the $^6$Li(p,$^3$He)$^4$He reaction at astrophysical energies, Few-Body systems {\bf 65}, 49 (2024).  
		\bibitem{GarridoGomez}
		L. Garrido-G\'omez, A. Vegas-D\'iaz, J. P. Fern\'andez-Garc\'ia,  M. A. G. Alvarez,  Systematic optical potentials for reactions with cluster-structured nuclei, Phys. Rev. C {\bf 109}, 054608 (2025).
		\bibitem{Wheeler}
		J. A. Wheeler, On the mathematical  description of light nuclei by the method of resonating group, Phys. Rev. {\bf 52}, 1107 (1937).
		\bibitem{Wildermuth}
		K. Wildermuth, Y. C. Tang, {\it  A unified theory of the nucleus} (Academic Press, New York, San Francisco, London, 1977). 
		\bibitem{Hiura}
		J. Hiura, I. Shimodaya, Alpha-particle model for Be$^{9}$, Prog. Theor. Phys. {\bf 30}, 585 (1963).
		\bibitem{Saito}
		S. Saito, S. Okai, P. Tamagaki, On he inner repuslive effect due to he Pauli principle between nuclear composite particles, Prog. Theor. Phys. {\bf 50}, 1561 (1973).
		\bibitem{Descovemont}
		P. Descovemont, Microscopic models for nuclear rates, J. Phys. G: Nucl. Part. Phys. {\bf 19}, S141 (1993).
		\bibitem{Fujiwara}
		Y. Fujiwara, Y. C. Tang, Reaction cross sections in light nuclear systems wih the multiconfiguration resonating-group method,  Prog. Theor. Phys. {\bf 93}, 711 (1995).
		\bibitem{Lashko}
		Yu. A. Lashko, V. S. Vasilevsky, V. I. Zhaba, Many-channel microscopic theory of resonance states and scattering processes in $^{9}Be$ and $^{9}B$, Phys. Rev. C {\bf 109}, 045803 (2024). 
		\bibitem{RGMNCSM}
		P. Navr\'atil, S. Quaglioni, R. Roth, W. Horiuchi, {\it Ab initio} calculaions of light-ion reactions, Prog. Theor. Phys. Suppl. {\bf 196}, 117 (2012).
		\bibitem{CDCC}
		M. Kawai, Formalism of the method of coupled discretized continuum channels, Prog. Theor. Phys. Suppl. {\bf 89}, 11 (1986).
		\bibitem{Yahiro}
		M. Yahiro, K. Ogata, T. Matsumoto, K. Minomo, The continuum dicretized coupled-channels method and its applications, Prog. Theor. Exp. Phys. {\bf 2012}, 01A206 (2012).
		\bibitem{Souza}
		F. A. Souza, A. Szanto de Toledo, N. Carlin, R. Liguori Neto, A. A. Suaide, M. M. Moura, E. M. Szano, M. G. Munhoz, J. Takahashi, C. Beck, S. J. Sanders,  Fusion of weakly bound light nuclei, Nucl. Phys. A {\bf 718}, 544c (2003).
		\bibitem{Guo}
		H. Guo,  Y. Watanabe, T. Matsumoto, K. Nagaoka, K. Ogata, M. Yahiro, Analysis of nucleon and trion emissions from nucleon-$^7$Li collisions below 20 MeV, Phys. Rev. C {\bf 99}, 034602 (2019). 
		\bibitem{Ichinkhorlo}
		D. Ichinkhorloo, Y. Hirabayashi, K. Kat$\bar{\text{o}}$, M. Aikawa, T. Matsumoto, S. Chiba, Analysis of $^{7}Li(n,n')^{7}Li^*$ reactions using he continuum-discretized 
		coupled-channels mehod, Phys. Rev. C {\bf 86}, 064604 (2012). 
		\bibitem{DiezTorez}
		A. Diez-Torres, I. J. Thompson, C. Beck, How does breakup influence the total fusion of $^{6,7}$Li at the Coulomb barrier? Phys. Rev. C {\bf 68}, 044607 (2003).
		\bibitem{Hiyama}
		E. Hiyama, Y. Kino, M. Kamimura, Gaussian expasion method for few-body systems, Prog. Part. Nucl. Phys. {\bf 51}, 223 (2003).
		\bibitem{Ogawa}
		Sh. Ogawa, K. Yoshida, Y. Chazono, K. Ogata, Three-body analysis reveals the significant contribution of minor $^5$He $s$-wave component in $^6$Li(p,2p)$^5$He cross section, Phys. Rev. C {\bf 111}, 034622 (2025).
		\bibitem{Hiyama97}
		E. Hiyama, M. Kamimura, T. Motoba, T. Yamada, Three- and four-body cluster models of hypernuclei using the $G$-matrix $\Lambda N$ interaction, Prog. Theor. Phys. {\bf 97}, 881 (1997). 
		\bibitem{Hiyama09}
		E. Hiyama, T. Yamada, Structure of light hypernuclei, Prog. Part. Nucl. Phys. {\bf 63}, 339 (2009).  
		\bibitem{EgorovnLi}
		M. V. Egorov, Inelastic scattering of fast neutrons on $^6$Li and  $^7$Li nuclei in three-body cluster model, Nucl. Phys. A {\bf 986}, 175 (2019).
		\bibitem{EgorovdLi}
		M. V. Egorov, V. I. Postnikov, Deuteron inelastic scattering on $^{6}$Li and $^7$Li nuclei within the three-body cluster model, Chinese Physics C {\bf 45}, 014108 (2021).
		\bibitem{Deltuva09}
		A. Deltuva, A. C. Fonseca, Three-body Faddeev-Alt-Grassberger-Sandhas approach to direct nuclear reactions, Phys. Rev. C {\bf 79}, 014606 (2009).
		\bibitem{EgorovFBSY}
		M. Egorov, Three-dimensional integral Faddeev equations without a certain symmetry, Few-Body Systems {\bf 66}, 24 (2025). 
		\bibitem{Forsyth}
		P. D. Forsyth, R. R. Perry, The $Li^7(He^3,\alpha)Li^6$ reaction between 1.3 and 5.5 MeV, Nucl. Phys. {\bf 67}, 517 (1965).	
		\bibitem{BonnNN}
		J. Haidenbauer, Y. Koike,  Separable representation of he Bonn nucleon-nucleon potential, Phys. Rev. C {\bf 33}, 439 (1986). 
		\bibitem{Gloeckle3N}
		W. Gl\"ockle, H. Wita\l a, D. H\"uber, H. Kamada, J. Golak, The three-nucleon continuum: achievements, challenges and applications,  Physics Reports {\bf 274}, 107 (1996). 	
		\bibitem{Adiobat}
		S. I., Vinitskii, L. I.   Ponomarev, The adiabatic representation in the three-body problem with the Coulomb interaction,  Sov. J. Part. Nucl. {\bf 13}, 557 (1982).
		\bibitem{Papp}
		B. K\'onya, G. L\'evai, Z. Papp, Continued fraction representation of the Coulomb Green's operator and unified description of bound, resonant and scattering states, Phys. Rev. C {\bf 61}, 034302 (2000).
		\bibitem{RubtsovaCoul}
		V. I. Kukulin,  O. A.  Rubtsova, Solving the scattering problem for charged particles by means of the packet discretization of the continuum, Theor. Math. Phys. {\bf 145}, 1711 (2005). 
		\bibitem{Oryu06}
		Sh. Oryu, Two- and three-charged-particle nuclear scattering in momentum space: A two-potential theory and a boundary condition model, Phys. Rev. C {\bf 73}, 054001 (2006). 
		\bibitem{Oryu16}
		Sh. Oryu, Y. Hiratsuka, T. Watanabe, Few-body problem in nuclear reactions {\it Beyond the horizon of the three-body Faddeev equations}, EPJ Web of Conferences {\bf 122}, 08001 (2016).
		\bibitem{MerkurCoul}
		S. P. Merkuriev, On the three-body coulomb scattering problem, Annals of Physics {\bf 130}, 395 (1980). 
		\bibitem{ENDF}
		Evaluated Nuclear Structure Data File (ENSDF) [electronic source] / National Nuclear Data Center, Brookhaven National Laboratory: http://www.nndc.bnl.gov/ensdf/.
		\bibitem{Youn}
		Li Ga Youn, G. M. Osetinskii, N. Sodnom, A. M. Govorov, I. V. Sizov, V. I. Salatski, Investigation of the $He^3+H^3$ reaction, Soviet Physics JETP {\bf 12}, 163 (1961).
		\bibitem{Kuhn}
		B. K\"uhn, B. Schlenk, Winkelverteilungen f\"ur die reaction $He^3+T$, Nucl. Phys. A {\bf 48}, 353 (1963).
		\bibitem{Moak}
		C. D. Moak, A study of the $H^3+He^3$ reactions, Phys. Rev. {\bf 92}, 383 (1953). 
		\bibitem{Smith}
		D. B. Smith, N. Jarmie, A. M. Lockett, $He^3+t$ reactions, Phys. Rev. {\bf 129}, 785 (1963)
		\bibitem{Nocken}
		U. Nocken, U. Quast, A. Richter, G. Schrieder, The reaction $^3H(^3He,d)^4He$ at very low energies: energy dependent violations of the Barshay-Temmer isospin theorem and highly excited states in $^6$Li, Nucl. Phys. A {\bf 213}, 97 (1973).
		\bibitem{Dwarakanath}
		M. R. Dwarakanath, H. Winkler, $^3\text{He}(^3\text{He},2p)^4\text{He}$ total cross-section measurement below the coulomb barrier, Phys. Rev. C {\bf 4}, 1532 (1971); $^3\text{He}(^3\text{He},2p)^4\text{He}$ and the termination of the proton-proton chain, Phys. Rev. C {\bf 9},  805 (1974). 
		\bibitem{Kudomi}
		N. Kudomi, M. Komori, K. Takahisa, S. Yoshida, Precise measurement of the cross section of $^3\text{He}(^3\text{He},2p)^4\text{He}$ by using $^3$He doubly charged beam, Phys. Rev. C {\bf 69}, 015802 (2004).
		\bibitem{Ming}
		W. Neng-Ming, B. N. Osetinskii, Ch. Nai-Kung, I. A. Chepushenko, Investigation of $^3$He+$^3$He reaction, Soviet J. of Nucl. Phys. {\bf 3}, 777 (1966).
		\bibitem{Brown}
		R. E. Brown, F. D. Correll, P. M. Hegland, J. A. Koepke, C. H. Poppe, $^3$He+$^3$He recation cross sections at 17.9,21.7 and 24.0 MeV, Phys. Rev. C {\bf 35}, 383 (1987). 
		\bibitem{Junker}
		M. Junker, A. D'Alessandro, S. Zavatarelli, C. Arpesella, E. Bellotti, C. Broggini, P. Corvisiero, G. Fiorentini, A. Fubini, G. Gervino, U. Greife,  C. Gustavino,  J. Lambert, P. Prati, W. S. Rodney, C. Rolfs, {\it et al.,} Cross section of $^3\text{He}(^3\text{He},2p)^4\text{He}$ measured at solar energies, Phys. Rev. C {\bf 57}, 2700 (1998).
		\bibitem{Krauss}
		A. Krauss, H. W. Becker, H. P. Trautvetter, C. Rolfs, Astrophysical $S(E)$ factor of $^3\text{He}(^3\text{He},2p)^4\text{He}$ at solar energies, Nucl. Phys. A {\bf 467}, 273 (1987).
		\bibitem{Bonetti}
		R. Bonetti, C. Broggini, L. Campajola, P. Corvisiero, A. D'Alessandro, M. Dessalvi, A. D'Onofrio, A. Fubini, G. Gervino, L. Gialanella, U. Greife, A. Guglielmetti, C. Gustavino, G. Imbriani, M. Junker, P. Prati {\it et al.,} First direct measurement of the $^3$He($^3$He,2p)$^4$He cross section within the Solar Gamov peak,  Phys. Rev. Lett. {\bf 82}, 5205 (1999). 
		\bibitem{Ludecke}
		H. L\"udecke, Tan Wan-Tjin, H. Werner, J. Zimmerer, The ractions $^6$Li($^3$He,$^3$He$_0$)$^6$Li, $^6$Li(d,d$_0$)$^6$Li, $^7$Li(d,d$_0$)$^7$Li and $^6$Li($^3$He,d$_{0,1}$)$^7$Be, Nucl. Phys. A {\bf 109}, 676 (1968).
		
		
		
	\end{thebibliography}
\end{document}